\documentclass[journal,twoside,web]{ieeecolor}
\pdfoutput=1
\usepackage{generic}
\usepackage{cite}
\usepackage{amsmath,amssymb,amsfonts}
\usepackage{algorithmic}
\usepackage{graphicx}
\usepackage{algorithm,algorithmic}
\usepackage{hyperref}
\hypersetup{hidelinks}
\usepackage{textcomp}
\usepackage{multirow}
\usepackage{makecell}

\usepackage{silence}
\newcommand{\etal}{\textit{et al. }}
\def\BibTeX{{\rm B\kern-.05em{\sc i\kern-.025em b}\kern-.08em
    T\kern-.1667em\lower.7ex\hbox{E}\kern-.125emX}}
\begin{document}
\title{The more, the better? Evaluating the role of EEG preprocessing for deep learning applications.}
\author{
Federico Del Pup$^{*}$, Andrea Zanola$^{*}$, Louis Fabrice Tshimanga, Alessandra Bertoldo and Manfredo Atzori
\thanks{This work was supported in part by the European Union’s Horizon Europe research and innovation programme under Grant agreement no 101137074 - HEREDITARY, in part by the STARS@UNIPD funding program of the University of Padova, Italy, through the project: MEDMAX.
\textit{(Federico Del Pup and Andrea Zanola are co-first authors.) (Corresponding author: Federico Del Pup)}}
\thanks{Federico Del Pup is with the Department of Information Engineering, University of Padova, Padova, 35131 Italy.
He is also with the Department of Neuroscience, University of Padova, Padova, 35121 Italy
and with the Padova Neuroscience Center, University of Padova, Padova, 35129, Italy (e-mail: federico.delpup@phd.unipd.it). }
\thanks{Andrea Zanola and Louis Fabrice Tshimanga are with the Padova Neuroscience Center, University of Padova, Padova, 35129 Italy,
and also with the Department of Neuroscience, University of Padova, Padova, 35121, Italy (e-mail: andrea.zanola@phd.unipd.it; louisfabrice.tshimanga@unipd.it).}
\thanks{Alessandra Bertoldo is with the Department of Information Engineering, University of Padova, Padova, 35131 Italy,
and also with the Padova Neuroscience Center, University of Padova, Padova, 35129 Italy (e-mail: alessandra.bertoldo@unipd.it). }
\thanks{Manfredo Atzori is with the Department of Neuroscience, University of Padova, Padova, 35121, Italy.
He is also with the Padova Neuroscience Center, University of Padova, Padova, 35129 Italy
and with the Information Systems Institute, University of Applied Sciences Western Switzerland (HES-SO Valais), Sierre, 3960 Switzerland (e-mail: manfredo.atzori@unipd.it).}
}

\maketitle

\begin{abstract}
%Background:
The last decade has witnessed a notable surge in deep learning applications for the analysis of electroencephalography (EEG) data, thanks to its demonstrated superiority over conventional statistical techniques.
%Motivation:
However, even deep learning models can underperform if trained with bad processed data.
While preprocessing is essential to the analysis of EEG data, there is a need of research examining its precise impact on model performance.
This causes uncertainty about whether and to what extent EEG data should be preprocessed in a deep learning scenario.
%Objective:
This study aims at investigating the role of EEG preprocessing in deep learning applications,
drafting guidelines for future research.
%improving current understanding.
%Methods:
It evaluates the impact of different levels of preprocessing, from raw and minimally filtered data to complex pipelines with automated artifact removal algorithms.
Six classification tasks (eye blinking, motor imagery, Parkinson's and Alzheimer's disease, sleep deprivation, and first episode psychosis) and four different architectures commonly used in the EEG domain were considered for the evaluation.
%Conclusion:
The analysis of 4800 different trainings revealed statistical differences between the preprocessing pipelines at the intra-task level, for each of the investigated models, and at the inter-task level, for the largest one.
Raw data generally leads to underperforming models, always ranking last in averaged score.
In addition, models seem to benefit more from minimal pipelines without artifact handling methods, suggesting that EEG artifacts may contribute to the performance of deep neural networks.
\end{abstract}


\section{Introduction}
\label{sec:introduction}
% ----- PUP ------
% BE CONCISE. BETTER TO GIVE MORE SPACE TO OTHER SECTIONS.

% What is EEG?
\IEEEPARstart{E}{LECTROENCEPHALOGRAPHY} (EEG) is a non-invasive functional imaging technique widely used to examine neural activity.
It is effectively employed in a multitude of domains, including Brain-Computer Interfaces (BCI)~\cite{BCIRew}, Emotion Recognition~\cite{EmotionRew}, Sleep Staging~\cite{SleepRw}, and the study of various diseases such as epilepsy~\cite{EpilepsyRew}, Parkinson's~\cite{ParkinsonRew}, and Alzheimer's~\cite{AlzheimerRew}.

% What brought deep learning to establish in this domain?
Due to their acquisition modality, EEG records have a low signal-to-noise ratio caused by a combination of biological and non-biological artifacts (e.g., blinking, subject movement, line noise, bad electrode placement) that can degrade the quality of the signal itself. 
Furthermore, neural activity is characterised by highly nonlinear patterns. 
Conventional statistical techniques and even classical machine learning approaches may be insufficient for capturing and modeling such complexity.

% Deep learning for EEG, benefits and consequences
To enhance the analysis of EEG data, Deep Learning (DL) has emerged as a new powerful resource in the domain \cite{DLReviewEEG}. 
As a subset of machine learning, deep learning focuses on training artificial neural networks, biologically inspired systems composed of multiple interconnected layers that perform both linear and nonlinear operations on the input data, typically a portion of a multichannel EEG recording. 
%As a subset of machine learning, deep learning focuses on training artificial neural networks, which are biologically inspired systems consisting of multiple interconnected layers. They perform both linear and nonlinear operations on the input data, which is usually a limited time-portion of a multichannel EEG records.

Such models have proven capable of discovering complex relationships by learning representations of data with multiple levels of abstraction~\cite{DLTrio}.
However, the increased learning capability is achieved at the cost of a higher computation and number of parameters to optimize. 
As a result, the demand for larger EEG repositories has increased significantly, leading to the release of numerous open-source datasets~\cite{EEGDataRew}. 
If, on the one hand, the growing number of repositories has facilitated the investigation of EEG-DL strategies, on the other hand, the time required to inspect and preprocess such data has increased remarkably.

% Focus on the preprocessing problem
% EEG data preprocessing is a complex and time-consuming phase composed of several steps; some of them (e.g., independent component rejection, bad channels or segments removal) still require supervision by an expert to be optimally performed.
EEG data preprocessing is a complex and time-consuming phase comprising several steps. Some of them, such as independent component rejection, bad channels or segments removal, still require expert supervision to be optimally performed.
Consequently, fully manual or semi-automated pipelines are unsuitable when dealing with large repositories with thousands of subjects.
Automatic pipelines, on the other hand, represent the optimal compromise between computational requirements and quality of the results.
For example, software libraries such as HAPPE~\cite{happe} and Automagic~\cite{automagic} can preprocess an entire dataset with a rich and customizable pipeline.
Others, like \textit{BIDSAlign}~\cite{BIDSalign}, further expand the preprocessing functionalities and allow to harmonise and combine multiple repositories, facilitating the investigation of novel DL strategies such as self-supervised learning~\cite{ssleeg}. 

% Intro to what we lack
EEG data preprocessing is extremely important and, if not done correctly, can lead to a significant drop in model performance, as thoroughly described in \autoref{sec:results}.
However, despite the large body of work published in the last decade, there is still no real consensus on the extent to which EEG data should be preprocessed for DL applications. 
% Different opinions on the matter
According to Roy \etal~\cite{Yannick}, who reviewed 154 works that applied DL to EEG, only a small percentage of researchers (23\%) preprocessed all the records with an extensive pipeline that included artifact handling, while the majority opted for a minimal preprocessing (47\%).
Furthermore, a small but still relevant percentage (15\%) fed raw data directly into the model, sometimes achieving similar or even better results compared to a preprocessing scenario~\cite{Aznan, Almogbel}.
Such numbers reflect the general idea that features could potentially be learned even with raw or minimally preprocessed data, reducing the need for domain-specific processing. 

The use of minimally preprocessed data can be justified if one assumes that noise and artifacts defined in one scenario could become important features in another one, leading to improvements in model accuracy, as stated in the recent analysis of Cui \etal on model interpretability~\cite{EEGInterp}.
In contrast, studies like~\cite{Hefron} found that a substantial preprocessing was required to prepare data for their specific task, leaving open the question of whether and to what extent EEG data preprocessing is necessary.
The first attempt to give a formal answer to this question was that of Kingphai \etal~\cite{PreproWL}, who investigated the effect of EEG preprocessing for the assessment of cognitive workload via DL approaches.
They found an improvement in model's performance as richer pipelines were applied to raw data.
However, it should be noted that their analysis focused on a single specific task and cannot be used as a general rule for other applications.

% Introducing our contribution
%Therefore, a broader and more detailed analysis is necessary and could be of great interest for the research community, who seek to fully exploit the capabilities of DL models for the analysis of EEG data.
Therefore, a more comprehensive investigation of the impact of preprocessing on DL applications to EEG data could be of great interest to the research community seeking to fully exploit the capabilities of DL models in this domain.

% alternative
%It would be beneficial for the research community to conduct a more comprehensive investigation into the impact of preprocessing on DL applications to EEG data. This could help to fully harness the potential of DL models in this field.

% Sentence removed
%Currently, the literature still lacks such an analysis, to the best of our knowledge.

% 2) OBJECTIVE OF THIS WORK
\textbf{Contributions}:
This work aims at thoroughly investigating the role of EEG preprocessing in DL applications, considering six classification tasks (eye blinking, motor imagery, Parkinson's and Alzheimer's disease, sleep deprivation, and first episode psychosis) and four DL architectures often used in the EEG domain.
First, it investigates the role of raw data and whether they can be effectively given to DL models without losing predictive power.
Then, it compares pipelines with different levels of complexity, from a minimal filtering to richer pipelines with established artifact handling automated algorithms.
Finally, it can also be useful to researchers who want to understand the role of EEG preprocessing on specific research areas.

% Next sections of the paper
\textbf{Paper structure}: The outline of this paper is as follows.
Section \ref{sec:methods} describes in detail the experimental setting.
Section \ref{sec:results} presents the results, which are discussed in \autoref{sec:discussion}.
Finally, a conclusion is drawn in \autoref{sec:conclusion}.

\section{Methods}
\label{sec:methods}
% ----- ANDREA E PUP ------

% ---------- ROY ET AL CHECKLIST (put XX if described)-----
% DATA
% XX   number of subjects
% XX   electrode montage
% XX   Shape of one example
% XX   Number of examples in training, validation, test
%
% EEG PREPROCESSING
% XX    Temporal filtering
% XX    Spatial filtering
% XX    Artifact handling techniques
% XX    Resampling
%
% NEURAL NETWORK ARCHITECTURE
% XX    Architecture type
% XX    Number of layers
% XX    Number of learnable parameters
%
% TRAINING HYPER
% XX    Parameter Initialization
% XX    Loss function
% XX    Batch size
% XX    Number of epochs
% XX    Stopping criterion
% XX    Regularization
% XX    Optimization algotrithm
% XX    Learning rate scheduler
% XX    Values of all hyperparameters (including random seed)
% XX    Hyperparameter search method
%
% PERFORMANCE AND MODEL COMPARISON
% XX    Performance metrics
% XX    validation scheme
% XX    Description of baseline models
%
% HARDWARE AND SOFTWARE (EXTRA)
%  XX   hardware
%  XX   software
%  XX   training time optimization (extra)

This section details the methodological design of the presented analysis.
It follows the checklist provided by Roy \etal~\cite{Yannick}, which covers all the information necessary to ensure results' reproducibility, namely: dataset selection, data preprocessing, models architecture, data partitioning, training hyperparameters, performance evaluation, and statistical analysis.

\subsection{Dataset Selection} %andrea
\label{data-sel}
The inclusion criteria used for the dataset selection are:
\begin{itemize}
    \item \textit{open access}: all dataset must be publicly available at their respective OpenNeuro's~\cite{OpenNeuro} webpage.
    \item \textit{raw data}: data should be raw, in order to allow applying progressively more preprocessing steps.
    \item \textit{number of channels}: the minimum was set to 19 channels plus A1-A2 for contralateral referencing, following American Clinical Neurophysiology Society (ACNS) guidelines~\cite{eeguide}. 
    %They recommend a minimum of 21 electrodes for clinical electroencephalography (19 plus A1-A2 for contralateral referencing).
\end{itemize}
Additionally, the datasets were selected in order to represent several EEG tasks:
\begin{itemize}
    \item `Eye': physiological classification of eyes open and eyes closed recordings.
    \item `MMI': motor movement imagery, a widely studied BCI application~\cite{MMIRew}.
    \item `Parkinson', `FEP', `Alzheimer': two and three classes pathology classification focused on relevant medical use-cases, i.e. Parkinson's, First Episode Psychosis and Alzheimer's diseases.
    \item `Sleep': normal sleep vs sleep deprivation recognition.
\end{itemize}
The list of datasets and tasks used is summarised in \autoref{tab: datasets}.
The next paragraphs are dedicated to a brief description of the datasets used.

\subsubsection{Eye - ds004148} %eye
% name, n subj, age
This dataset named ``A test-retest resting and cognitive state EEG dataset"~\cite{ds004148}, contains resting (eyes closed, eyes open) and cognitive state (subtraction, music, memory) recordings for 60 young healthy subjects (age $20.0 \pm 1.9$ years).
% structure
All the subjects have only one session (duration $300.0 \pm 0.0$ seconds) in which all the tasks are performed.
% selection 
This dataset is used to perform the `Eye' task, thus only two recordings have been considered. 

\subsubsection{MMI - ds004362} %mmi
% name, n subj, age
This dataset named ``EEG Motor Movement/Imagery Dataset" (EEGMMI)~\cite{ds004504}, contains recordings for rest and both imagery and physical movements for 109 adult healthy subjects (age $39 \pm 11$ years).
% structure
All subjects have one session (duration $114.5 \pm 22.0$ seconds) with 14 tasks: two are rest (eyes open/closed), and the other twelve are three repetitions of four tasks.
\begin{itemize}
    \item Task 1: open and close left or right fist;
    \item Task 2: imagine opening and closing left or right fist;
    \item Task 3: open and close both fists or feet;
    \item Task 4: imagine opening and closing both fists or feet;
\end{itemize}
% selection 
This dataset is used to perform the `MMI' task, thus only Task 2 is considered.
Subjects with ID 88, 92 and 100, have been removed from the analysis since they have a different sampling rate and inconsistent trial length with respect to the others, as done in other works~\cite{mmi_rm1},~\cite{mmi_rm2}.

\subsubsection{Parkinson's - ds002778} %pds
% name, n subj, age
This dataset named ``UC San Diego Resting State EEG Data from Patients with Parkinson's Disease"~\cite{ds002778}, contains rest eyes open EEG recordings (duration $195.7 \pm 18.8$ seconds) from 12 Parkinson's patients and 16 age matched healthy controls (respectively $63.3 \pm 8.2$ years, $63.5 \pm 9.7$ years.)
% structure
Healthy subjects have only one session, instead Parkinson's patients have two sessions: one named `ses-off' containing recordings where patients discontinued medication use at least 12 hours before the session; one named `ses-on', where patients are under medication. 
% selection 
This dataset is used to perform the `control' vs `parkinson-off' task (named `Parkinson'), thus only `ses-off' was considered for the Parkinson's patients.

\subsubsection{Parkinson's - ds003490} %pds
% name, n subj, age
This dataset, named ``EEG: 3-Stim Auditory Oddball and Rest in Parkinson's"~\cite{ds003490}, contains rest eyes open/closed and auditory oddball EEG recordings (duration $595.9 \pm 74.0$ seconds) from 25 Parkinson's patients and 25 age matched healthy controls (respectively $69.7 \pm 8.7$ years, $69.3\pm 9.6$ years).
% structure
Healthy subjects have only one session, while Parkinson's patients have two sessions: one in which patients discontinued medication use at least 15 hours before the session and the other where patients are under medication.
% selection 
This dataset is used to perform the `control' vs `parkinson-off' task (named `Parkinson'), thus only one session was considered for the Parkinson's patients, together with data from dataset ds002778.

\subsubsection{Alzheimer's - ds004504} %alz
% name, n subj, age
This dataset named ``A dataset of EEG recordings from: Alzheimer's disease, Frontotemporal dementia and Healthy subjects"~\cite{ds004504}, contains EEG resting state eyes closed recordings (duration $802.2 \pm 140.3$ seconds) from 36 Alzheimer's patients, 23 subjects with diagnosed frontotemporal dementia and 29 age matched healthy controls (respectively $66.4 \pm 7.9$ years, $63.7\pm8.2$ years and $67.9\pm5.4$ years).
% structure
%All the subjects have only one session.
% selection 
This dataset is used to perform the `control' vs `alzheimer' vs `frontotemporal dementia' task (named `Alzheimer').

\subsubsection{Sleep - ds004902} %sleep
% name, n subj, age
This dataset named ``A Resting-state EEG Dataset for Sleep Deprivation"~\cite{ds004902}, contains EEG resting state eyes open recordings from 71 adult healthy subjects (age $20.0 \pm 1.4$ years).
% structure
All the subjects have two sessions (duration $295.5 \pm 26.7$ seconds), one for sleep deprivation and the other for normal sleep state. 
Only some subjects have eyes closed recording. For this reason, only eyes open records are used. 
% selection 
This dataset is used to perform the `normal' vs `sleep deprived' task (named `Sleep').

\subsubsection{FEP - ds003947} %psy
% name, n subj, age
This dataset named ``EEG: First Episode Psychosis vs. Control Resting Task 2"~\cite{ds003947}, contains EEG resting state eyes open recordings (duration $310.8 \pm 14.4$ seconds) from 31 first episode schizophrenia (FESz) subjects and 30 age matched healthy controls (respectively $23.4 \pm 4.7$ years, $24.2 \pm 5.1$ years).
% structure
%All the subjects have only one session.
% selection 
This dataset is used to perform the `control' vs `first episode psychosis' task (named `FEP').

%\subsubsection{ds004584} %cgn
% name, n subj, age
%This dataset named ``Evoked mid-frontal activity predicts cognitive dysfunction in Parkinson's disease" \cite{ds004584}, contains EEG resting state eyes open recordings from 100 Parkinson's subjects and 49 healthy controls.
%These two groups are age-matched, where for the Parkinson's group the mean age is M=68.5, SD=8.1 years, while for the control group is M=70.9, SD=7.6 years.
%Parkinson's patients can be divided in Parkinson's with dementia (PDD, 19 subjects), with mild cognitive impairment (PDMCI, 34 subjects) and normal Parkinson (PD, 47 subjects). 
%This subdivision was done by the authors of the dataset with the following Montreal Cognitive Assessment (MOCA) cut-offs: for PDD, MOCA$<22$, while for PD, MOCA$\geq$26.
% structure
% All the subjects have only one session.
% selection 
% This dataset will be used to perform the `control' vs `parkinson' task (named `cgn').

\subsection{Data Preprocessing}%andrea
\label{data-prep}
% Descrizione dei vari step di preprocessing e le relative pipeline scelte, con tabella riepilogativa. Nel dettaglio e giustificando ogni scelta e i software usati.

All the recordings selected from the datasets presented in \autoref{data-sel} have been preprocessed with four pipelines composed by the most common EEG-preprocessing steps used by both the medical~\cite{happe, automagic} and the deep learning community~\cite{Yannick}.
The preprocessing steps are selected both on the basis of the different levels of complexity/computation required, and on the impact on data.
%Moreover these four pipelines are composed by the most common EEG-preprocessing steps used by both the medical \cite{happe, automagic} and the deep learning community \cite{Yannick}.
Preprocessing was done with \textit{BIDSAlign}\footnote{https://github.com/MedMaxLab/BIDSAlign}~\cite{BIDSalign}, a MATLAB\textsuperscript{\textregistered}-based library developed on top of EEGLAB~\cite{eeglab}, one of the most used frameworks for EEG preprocessing.
The four pipelines, `Raw', `Filt', `ICA' and `ICA+ASR', and their preprocessing steps, are summarized in \autoref{tab: pipelines}, while the preprocessing steps and  parameters are described below.

\renewcommand{\arraystretch}{1.6} % Default value: 1
\begin{table}[!t]
    \centering
    \caption{Preprocessing pipelines considered in this work.}
    \begin{tabular}{ccccc}
        \Xhline{2\arrayrulewidth} \\[-1,3em]
        preprocessing step & Raw & Filt & ICA & ICA+ASR\\
        \\[-1,3em] \Xhline{2\arrayrulewidth}
        \multicolumn{1}{c|}{non-EEG channels removal} & \multicolumn{1}{c|}{\checkmark} & \multicolumn{1}{c|}{\checkmark} & \multicolumn{1}{c|}{\checkmark} & \checkmark \\
        \hline
        \multicolumn{1}{c|}{time segments removal} & \multicolumn{1}{c|}{} & \multicolumn{1}{c|}{} & \multicolumn{1}{c|}{\checkmark} & \checkmark \\
        \hline
        \multicolumn{1}{c|}{baseline removal} & \multicolumn{1}{c|}{} & \multicolumn{1}{c|}{\checkmark} &  \multicolumn{1}{c|}{\checkmark} &  \checkmark \\
        \hline
        \multicolumn{1}{c|}{resampling} & \multicolumn{1}{c|}{\checkmark} & \multicolumn{1}{c|}{\checkmark} & \multicolumn{1}{c|}{\checkmark} & \checkmark \\
        \hline
        \multicolumn{1}{c|}{filtering} & \multicolumn{1}{c|}{} & \multicolumn{1}{c|}{\checkmark} & \multicolumn{1}{c|}{\checkmark} & \checkmark \\
        \hline
        \multicolumn{1}{c|}{independent component analysis} & \multicolumn{1}{c|}{} & \multicolumn{1}{c|}{} & \multicolumn{1}{c|}{\checkmark} & \checkmark \\
        \hline
        \multicolumn{1}{c|}{automatic component rejection} & \multicolumn{1}{c|}{} & \multicolumn{1}{c|}{} & \multicolumn{1}{c|}{\checkmark} &  \checkmark \\
        \hline
        \multicolumn{1}{c|}{bad-channel removal} & \multicolumn{1}{c|}{} & \multicolumn{1}{c|}{} & \multicolumn{1}{c|}{} & \checkmark \\
        \hline
        \multicolumn{1}{c|}{bad-time windows correction (ASR)} & \multicolumn{1}{c|}{} & \multicolumn{1}{c|}{} & \multicolumn{1}{c|}{} & \checkmark \\
        \hline
        \multicolumn{1}{c|}{spherical interpolation}  & \multicolumn{1}{c|}{} & \multicolumn{1}{c|}{} & \multicolumn{1}{c|}{} & \checkmark \\
        \hline 
        \multicolumn{1}{c|}{re-reference} & \multicolumn{1}{c|}{} & \multicolumn{1}{c|}{\checkmark} & \multicolumn{1}{c|}{\checkmark} & \checkmark \\
        \hline
        \multicolumn{1}{c|}{template alignment} & \multicolumn{1}{c|}{\checkmark} & \multicolumn{1}{c|}{\checkmark} & \multicolumn{1}{c|}{\checkmark} & \checkmark \\
        \hline
    \end{tabular}
    \label{tab: pipelines}
\end{table}
\renewcommand{\arraystretch}{1} % Default value: 1

%The preprocessing steps and corresponding parameters are described below:
\begin{itemize}
    \item \textit{non-EEG channels removal}: all non-EEG data such as electrocardiogram (ECG) or electrooculogram (EOG) included as extra channels are removed at the beginning.
    \item \textit{time segments removal}: the first and last 8 seconds are removed.  This avoids potential initial and final divergences in the recording due to technical issues like switching on-off the EEG amplifier.
    \item \textit{baseline removal}: DC-component is removed by subtracting the channel-wise mean from each data point.
    \item \textit{resampling}: the resampling frequency was set to 250Hz.
    \item \textit{filtering}: EEG records were filtered with a pass-band Hamming windowed sinc FIR filter between 1Hz and 45Hz. This avoids the use of a notch filter for line artifacts and allows for the removal of low frequency artifacts.
    \item \textit{independent component analysis}: using the `\textit{runica}' algorithm~\cite{infomaxica}, with no limit to the number of components that can be found.
    \item \textit{automatic component rejection}: independent components were rejected automatically using IClabel~\cite{ICLabel}, with thresholds for the brain class set between 0 and 0.1, while for the other six artifacts it was set between 0.9 to 1.
    This is a good trade-off between discarding lot of components, and keeping what could really be brain signal.
    \item \textit{bad-channel removal}: flat or extremely noisy channels are removed with the EEGLAB's `\textit{clean\_rawdata}' plugin, using default parameters.
    \item \textit{bad-time windows removal}: this step was performed with Artefact Subspace Reconstruction (ASR) algorithm~\cite{asr}, using default parameters; bad-time windows are corrected and not removed.
    \item \textit{spherical interpolation}: previously automated removed channels are interpolated with the spherical method~\cite{Spherical}.
    \item \textit{re-reference}: records are re-referenced to the common average.
    \item \textit{template alignment}: to the International Federation of Clinical Neurophysiology (IFCN) 10-10 standard containing 61 channel. 
    This step allows the alignment of EEG recording acquired with different electrodes and channel systems to the same expected input of a DL architecture.
    Consequently, it makes possible the merging of the ds002778 and ds003490 datasets for the Parkinson's task. 
\end{itemize}

%An additional note must be said for the dataset ds004362 (EEGMMI - motor imagery task).
%Firstly, because the original sampling rate was 160 Hz, which is lower than the chosen 250 Hz resampling value, no resampling was done during the initial preprocessing with \textit{BIDSAlign} to avoid up-sampling.
%Secondly, the time segment removal step was skipped to maintain as many trials as possible. Instead, a specific trial extraction step was done at the end of the preprocessing.
For dataset ds004362 (EEGMMI - motor imagery task), whose records have a sampling rate of 160 Hz, resampling to 250Hz was not performed to avoid upsampling.
In addition, a dedicated trial extraction step was done in place of time segment removal to preserve as many trials as possible. 

To conclude, BIDSAlign preprocessed data went through two additional preparatory steps for the deep learning part.
%To conclude, two additional preparatory steps were included for the deep learning part.
First, EEG records were further downsampled to 125 Hz to reduce the sample size and improve computational resources.
Second, each EEG record was normalized by applying the \textit{z}-score operator along the EEG channel dimension, i.e., by transforming each channels' signal so to get $\mu\text{=0}$ and $\sigma\text{=1}$.
Data normalization have proven to be beneficial for EEG-DL applications~\cite{datanorm}.

\subsection{Implementation Details} %pup
\label{implem-details}
The remaining steps of the experiment were implemented inside a Python environment.
In particular, models were trained using \textit{SelfEEG}\footnote{https://github.com/MedMaxLab/selfEEG}~\cite{Selfeeg}, a comprehensive library for EEG-DL applications built on top of \textit{Pytorch}~\cite{pytorch}; statistical analysis relied on the \textit{Scipy} framework~\cite{scipy}, and figures were generated with \textit{Seaborn}~\cite{seaborn} coupled with \textit{statannotations}~\cite{statann}.
%Results are visualized with the \textit{Seaborn}'s colorblind palette.
Experiments were conducted on two NVIDIA Tesla V100 GPU devices to allow multiple models to be trained in parallel, speeding up the entire process.
The maximum GPU memory allocation was 6.18 GB.
Additional implementation choices not reported in the following subparagraphs can be found in subsection I-A of the supplementary materials.

\subsubsection{EEG Architectures} %pup
\label{model-selc}

Different models are expected to give different results.
While this aspect is desired in other works that aim to compare novel strategies against baseline methods, here it can be considered as an external factor that could affect the results of the statistical analysis.
For this reason, this work utilizes four of the most frequently used architectures , namely: EEGNet~\cite{eegnet}, ShallowConvNet~\cite{shallow}, DeepConvNet~\cite{shallow}, and FBCNet~\cite{fbcnet}.

Models have been chosen considering several factors, such as their overall usage among the research community and their computational requirements (e.g., number of layers/parameters, total training time, GPU memory allocation).
All models are composed of a series of layers, including among the others convolutions with horizontal or vertical kernels that compute operations respectively on the temporal (EEG samples) and on the spatial (EEG channels) dimension.
Furthermore, FBCNet incorporates a filter bank layer, a concept proven successful in the BCI domain. 
This layer is particularly suitable for this study, as certain artifact handling steps, such as the IC rejection, significantly influence the signal's power spectral density within specific frequency bands (e.g., delta band [0.1-4] Hz, lower-gamma band [30-45] Hz).
Additional information and a full description of each model can be found in the relative papers.

\subsubsection{Data Partition} %pup
\label{data-part}

EEGs are known to be characterized by a high inter-subject variability~\cite{SubjAwSSL}. 
The training could end up producing a model whose embeddings are dominated by subject-specific characteristics that could lead to misleading results: extremely good on subjects already seen but rather poor on unseen ones (see subsection A.3 of the supplementary material).
However, most of the real-world applications expect models to be used on new subjects, which is why recent works stressed the importance of inter-subject evaluations, such as the Leave-N-Subject-Out (LNSO) or the Leave-One-Subject-Out (LOSO) in the limit case of $\text{N=1}$~\cite{LOSO}.
The LNSO is a cross validation procedure where folds are created by partitioning the dataset at the subject level, i.e., by assigning all the subject's samples to the same set (training or validation).
This procedure provides a subject-independent estimate of the performance for new subjects and better reflects real-world applications.

\begin{figure}[!t]
    \centering
    % [trim={left bottom right top},clip]
    % \includegraphics[trim={1cm 4cm 6cm 2cm}, width=\linewidth]{Images/NestedKfold3.pdf}
    \includegraphics[trim={1cm 5,5cm 5cm 2cm}, width=0.95\linewidth]{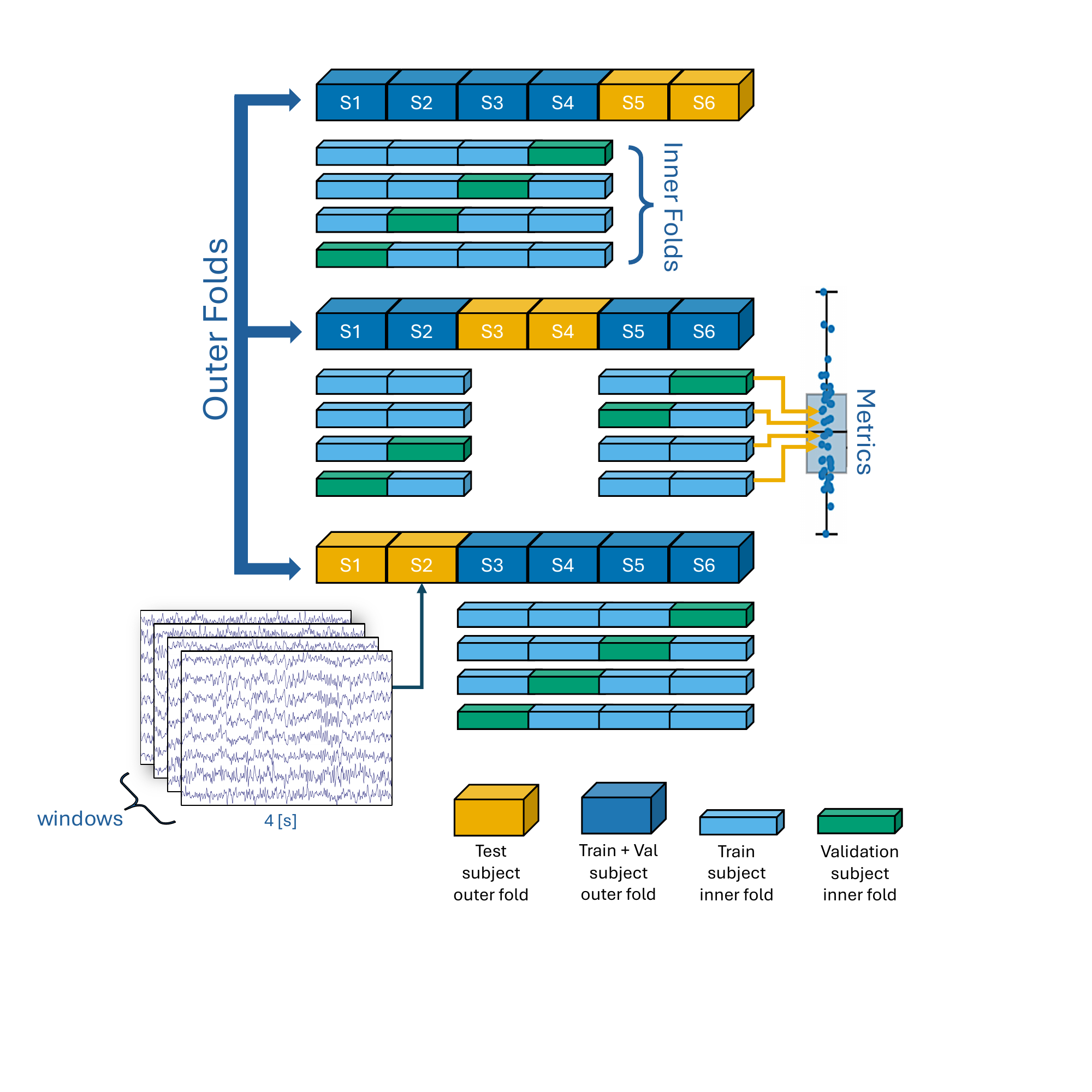}
    \caption{
    Schematization of the Nested Leave-N-Subject-Out Cross Validation.
    Each outer fold is used to determine the subjects to use as test set.
    Then, the remaining subjects are partitioned multiple times in training and validation sets based on the decided number of inner folds.
    A model is trained for each of the inner folds and its performance is evaluated on the test set of the corresponding outer fold.
    The procedure is repeated for each outer fold, resulting in a total of $N_{\text{outer}} \times N_{\text{inner}}$ training, leaving to an unbiased estimate of the centrality measure used for the statistical analysis.
    All subject's time windows, of length 4 seconds, are consequently put together in one of the three split sets.
    }
    \label{fig: nested_kfold}
\end{figure}

Further extending the LNSO strategy in a training-validation-test scenario, better suited for DL applications, this work proposed a Nested Leave-N-Subject-Out (N-LNSO) cross validation, which is schematized in \autoref{fig: nested_kfold}.
In short, the N-LNSO can be considered as a combination of two nested LNSO.
The first LNSO determines a set of outer folds, used to assign the subjects to use as test set; then, the remaining subjects are partitioned multiple times into training and validation sets based on the second LNSO, which determines the number of inner folds.
A model is trained for each of the inner folds and its performance is evaluated on the test set of the corresponding outer fold.
The procedure is repeated for each outer fold, resulting in a total of $N_{\text{outer}} \times N_{\text{inner}}$ training.
By doing so, each subject is guaranteed to participate at least once in every split set (training, validation, and test).

It is important to stress that, while a classical nested k-fold is used to nest the hyperparameters optimization procedure under the model selection procedure, the goal of the N-LNSO is to better estimate the model performance distribution against unseen subjects and to provide an unbiased measure of centrality (with the test set of unseen subjects) for the following statistical analysis~\cite{overfitting}.

For this work, the number of outer folds is $N_{outer}=10$ and inner folds is $N_{inner}=5$.
Considering also the number of preprocessing pipelines $N_{pipeline}=4$, models $N_{model}=4$ and tasks $N_{task}=6$, the total number of training was:
\begin{equation}
    \label{eq: tot_number}
        N_{pipeline}\times N_{model}\times N_{task}\times N_{outer} \times N_{inner} = 4800
\end{equation}
Splits were stratified in order to preserve class ratios.

Once assigned to a specific set, EEG records were divided in windows of 4 seconds without overlap (4.1 seconds for MMI to preserve the original length).
The final portion of the recording was discarded if not long enough to generate an additional sample.
Consequently, considering that the resampling and template alignment steps were included in each preprocessing pipeline, the input dimension was 61x500 (61x512 for the motor imagery).
Additional information for each dataset (task) and its total number of samples are reported in \autoref{tab: datasets}.

\renewcommand{\arraystretch}{1.6} % Default value: 1
\begin{table*}[!ht]
\centering
\caption{Dataset And Tasks Selected For This Work.}
\begin{tabular}{cccccccc}
\Xhline{2\arrayrulewidth} \\[-1,3em]
dataset ID & \makecell{original \\ reference} & \makecell{original \\ number \\ of channels} & \makecell{original\\ sampling \\ rate [Hz]} & \makecell{number of\\ subjects} & \makecell{number \\of windows\\raw/filt (ica/ica+asr)$^{*}$} & \makecell{task\\ description} & \makecell{task\\ acronym} \\
\\[-1,3em] \Xhline{2\arrayrulewidth}
\multicolumn{1}{c|}{ds004148} & \multicolumn{1}{c|}{FCZ} & \multicolumn{1}{c|}{64} & \multicolumn{1}{c|}{500} & \multicolumn{1}{c|}{60} & \multicolumn{1}{c|}{9000 (8520)} & \multicolumn{1}{c|}{Eyes Open vs Eyes closed} & Eye \\ \hline
\multicolumn{1}{c|}{ds004362} & \multicolumn{1}{c|}{A1-A2} & \multicolumn{1}{c|}{64} & \multicolumn{1}{c|}{160} & \multicolumn{1}{c|}{109$^{**}$} & \multicolumn{1}{c|}{4748 (4748)} & \multicolumn{1}{c|}{Imagine open/close left vs right fist} & MMI \\
\hline
\multicolumn{1}{c|}{ds002778} & \multicolumn{1}{c|}{CMS-DRL} & \multicolumn{1}{c|}{41} & \multicolumn{1}{c|}{512} & \multicolumn{1}{c|}{31} & \multicolumn{1}{c|}{\multirow{2}{*}{8932 (8608)}} & \multicolumn{1}{c|}{\multirow{2}{*}{Control vs Parkinson's off-medication}} & \multirow{2}{*}{Parkinson} \\ \cline{1-5}
\multicolumn{1}{c|}{ds003490} & \multicolumn{1}{c|}{CPZ} & \multicolumn{1}{c|}{64} & \multicolumn{1}{c|}{500} & \multicolumn{1}{c|}{50} & \multicolumn{1}{c|}{} & \multicolumn{1}{c|}{} &  \\
\hline
\multicolumn{1}{c|}{ds004504} & \multicolumn{1}{c|}{A1-A2} & \multicolumn{1}{c|}{19} & \multicolumn{1}{c|}{500} & \multicolumn{1}{c|}{88} & \multicolumn{1}{c|}{17604 (17252)} & \multicolumn{1}{c|}{Control vs Alzheimer vs FT-dementia} & Alzheimer \\
\hline
\multicolumn{1}{c|}{ds004902} & \multicolumn{1}{c|}{FCZ} & \multicolumn{1}{c|}{61} & \multicolumn{1}{c|}{500} & \multicolumn{1}{c|}{71} & \multicolumn{1}{c|}{10482 (9914)} & \multicolumn{1}{c|}{Normal vs Sleep deprived} & Sleep \\
\hline
\multicolumn{1}{c|}{ds003947} & \multicolumn{1}{c|}{TP9} & \multicolumn{1}{c|}{61} & \multicolumn{1}{c|}{1000} & \multicolumn{1}{c|}{61} & \multicolumn{1}{c|}{4718 (4474)} & \multicolumn{1}{c|}{Control vs First Episode Psychosis (FEP)} & FEP \\ 
\hline
\multicolumn{7}{l}{${ }^{\text{*ICA and ICA+ASR pipelines have fewer samples due to the segment removal step,}}$ ${ }^{\text{**three subjects were excluded from the analysis}}$}
\end{tabular}
\label{tab: datasets}
\end{table*}
\renewcommand{\arraystretch}{1} % Default value: 1

\subsubsection{Training Hyperparameters} %pup
\label{train-hyp}

The whole experiment was conducted using a custom seed (83136297) to improve the reproducibility of the results.
Each model was initialized using Pytorch's default settings, which are specific for each type of layer.
Model architecture was the same as proposed by original authors except for the max norm constraints, which were not used in this work.
Models were trained using Adam optimizer ($\beta_{1} = 0.9$, $\beta_{2} = 0.999$, no weight decay)~\cite{adam}, batch size of 64, cross entropy as loss function, and an exponential learning rate scheduler ($\gamma = 0.995$).
The maximum number of epochs was set to 100, coupled with an early stopping with patience of 25 epochs (monitoring on the validation loss).
Since different models are characterized by different learning curves, which can also vary depending on the investigated task, a custom learning rate was chosen for each possible couple (task, model), keeping it equal among all the pipelines.
Learning rates were selected by evaluating the performance of a subset of models on the validation set in terms of consistency (median and overall rank of different metrics), searching on a discrete grid of 13 possible values: $1.0\cdot10^{-3}$, $7.5\cdot10^{-4}$, $5.0\cdot10^{-4}$, $2.5\cdot10^{-4}$, $1.0\cdot10^{-5}$, $7.5\cdot10^{-5}$, $5.0\cdot10^{-5}$, $2.5\cdot10^{-5}$, $1.0\cdot10^{-6}$.
The list of used learning rate is reported in \autoref{tab: learningrate}.

\renewcommand{\arraystretch}{1.6} % Default value: 1
\begin{table}[!t]
\caption{Learning rate grid.}
\begin{tabular}{ccccc}
\Xhline{2\arrayrulewidth} \\[-1,3em]
Task & EEGNet & ShallowNet & DeepConvNet & FBCNet \\
\\[-1,3em] \Xhline{2\arrayrulewidth}
\multicolumn{1}{c|}{Eye} & \multicolumn{1}{c|}{$5.0\cdot10^{-4}$} & \multicolumn{1}{c|}{$1.0\cdot10^{-3}$} & \multicolumn{1}{c|}{$7.5\cdot10^{-4}$} & $7.5\cdot10^{-4}$ \\ \hline
\multicolumn{1}{c|}{MMI} & \multicolumn{1}{c|}{$1.0\cdot10^{-3}$} & \multicolumn{1}{c|}{$7.5\cdot10^{-4}$} & \multicolumn{1}{c|}{$7.5\cdot10^{-4}$} & $1.0\cdot10^{-3}$ \\ \hline
\multicolumn{1}{c|}{Parkinson} & \multicolumn{1}{c|}{$1.0\cdot10^{-4}$} & \multicolumn{1}{c|}{$2.5\cdot10^{-4}$} & \multicolumn{1}{c|}{$2.5\cdot10^{-4}$} & $2.5\cdot10^{-4}$ \\ \hline
\multicolumn{1}{c|}{Alzheimer} & \multicolumn{1}{c|}{$7.5\cdot10^{-4}$} & \multicolumn{1}{c|}{$5.0\cdot10^{-5}$} & \multicolumn{1}{c|}{$7.5\cdot10^{-4}$} & $7.5\cdot10^{-5}$ \\ \hline
\multicolumn{1}{c|}{Sleep} & \multicolumn{1}{c|}{$1.0\cdot10^{-3}$} & \multicolumn{1}{c|}{$5.0\cdot10^{-5}$} & \multicolumn{1}{c|}{$2.5\cdot10^{-4}$} & $1.0\cdot10^{-4}$ \\ \hline
\multicolumn{1}{c|}{FEP} & \multicolumn{1}{c|}{$1.0\cdot10^{-4}$} & \multicolumn{1}{c|}{$7.5\cdot10^{-5}$} & \multicolumn{1}{c|}{$1.0\cdot10^{-3}$} & $1.0\cdot10^{-5}$ \\ \hline
\end{tabular}
\label{tab: learningrate}
\end{table}
\renewcommand{\arraystretch}{1} % Default value: 1

\subsubsection{Performance Evaluation} %pup
\label{performance}
The performance of each model was evaluated using the balanced accuracy, which is preferred over the unbalanced one in biomedical applications as it takes into account differences in class ratio.
By definition, balanced accuracy is the macro average of recall scores per class, so for a multi-class classification problem, with $C$ classes:
\begin{equation}
    \text{Accuracy}_{\text{Balanced}} = \dfrac{1}{C}\sum_{i=1}^C\dfrac{\text{TP}_i}{\text{TP}_i + \text{FN}_i}
    \label{equ: BA_Nclasses}
\end{equation}
where $\text{TP}_i$ and $\text{FN}_i$ stands respectively for the true positive and the false negative for class $i$.

Other metrics, such as the weighted versions of recall, precision, and F1 score, were evaluated as well. 
The statistical analysis against such metrics is reported in subsection I-B of the supplementary materials. 

\subsection{Statistical Analysis} %andrea
\label{statistic}
The main objective of this work is to evaluate if different amounts of preprocessing to the data can lead to different results in terms of metric under evaluation, in this case, the balanced accuracy.

The procedure of splitting the dataset into three partitions (train, validation and test) and applying the Nested Leave-N-Subject-Out strategy described in \autoref{data-part}, gives as result an ensemble of trained models.
Consequently, for each of the six tasks there are 200 trained models, with their respective evaluation, divided in groups of 50 (a single N-LNSO) for each of the four pipelines under investigation.

Considering one task (intra-task level), the distributions of the results for the four pipelines are compared pairwise with a Wilcoxon's signed rank test~\cite{wilcoxon}.
Since classification results of the same data, preprocessed with different pipelines are dependent, the Wilcoxon's test is preferred over the Mann-Whiteny U test~\cite{mannwhit}.
The p-values are corrected for multiple comparison using the Holm's~\cite{holm} procedure within the same task, and visualized together with the box-plots in \autoref{fig: models_results} (Panels A-D I).

Median balanced accuracy is used as measure of centrality for the distribution of the 50 models' performances; it is preferable over the sample mean, when distributions are not normal or skewed.
The Friedmann's non-parametric test~\cite{Friedman} is used to evaluate the differences on the medians between the four pipelines; in particular, it ranks the preprocessing pipeline for each dataset separately, the best performing pipeline getting rank 1, the second best rank 2, and so on.
This statistical test is well suited for such an analysis as described in~\cite{demsar} and~\cite{raschka2020model} and it was previously applied for analogous purposes in~\cite{effectprep}.
The core idea is to use different tasks, associated to one or more datasets, as evidence to demonstrate that pipelines have an impact in the classification result (inter-task level).
%Moreover the Friedmann's test is non-parametric, thus does not relies on the ANOVA's assumptions, which is its parametric counterpart.
%Therefore, it does not assumes that samples are drawn from a normal distributions and does not assumes sphericity \cite{demsar}.

Let $r_i^j$ be the rank of the $j$-th of $k$ pipelines on the $i$-th of $N$ task; the Friedman's test compares the average ranks of the $k$ pipelines.
\begin{equation}
    R_j = \frac{1}{N}\sum_{i=1}^{N} r_i^j
\end{equation}

Under the null-hypothesis $H_0$, which states that all the pipelines are equivalent, so their ranks $R_j$ should be equal; thus the Friedman statistic
\begin{equation}
    \chi_F^2 =\frac{12N}{k(k+1)}\sum_{j=1}^{k} R_j^2 - 3N(k+1)
    \label{equ: chi2F}
\end{equation}
is distributed according to a $\chi^2$ distribution, with $k-1$ degrees of freedom, when $N$ and $k$ are big enough.
For $k=4$ (pipelines) and $N=6$ (tasks), the Friedman's statistic at critical value $\alpha=0.05$ is $\chi_{approx}^2=7.815$ within the $\chi^2$ approximation, while the tabulated value is $\chi_{tabular}^2=7.6$.

If $H_0$ is rejected, a post-hoc test is needed to understand which pipelines are significantly different.
The Nemenyi's test~\cite{nemenyi} is used when all pipelines are compared to each other~\cite{demsar}.
The performances of two pipelines, $i$ and $j$, are significantly different if the corresponding average ranks $R_i$ and $R_j$ differ by at least the critical difference (CD):
\begin{equation}
    |R_i - R_j|\geq CD = q_\alpha \sqrt{\frac{k(k+1)}{6N}}
    \label{equ: CD}
\end{equation}
where critical values $q_\alpha$ are based on the Studentized range statistic divided by $\sqrt{2}$. 
A table with critical values can be found in~\cite{demsar}; for $k=4$ (pipelines) and $N=6$ (tasks), the critical distance is $CD=1.915$.
Results are visualized using the critical difference diagram, in \autoref{fig: models_results} (Panels A-D II).

The described procedure is repeated using four different deep neural networks architectures, as described in \autoref{model-selc}.

\section{Results}
\label{sec:results}
\begin{figure*}[!t]
    \centering
    \includegraphics[width=0.47\textwidth]{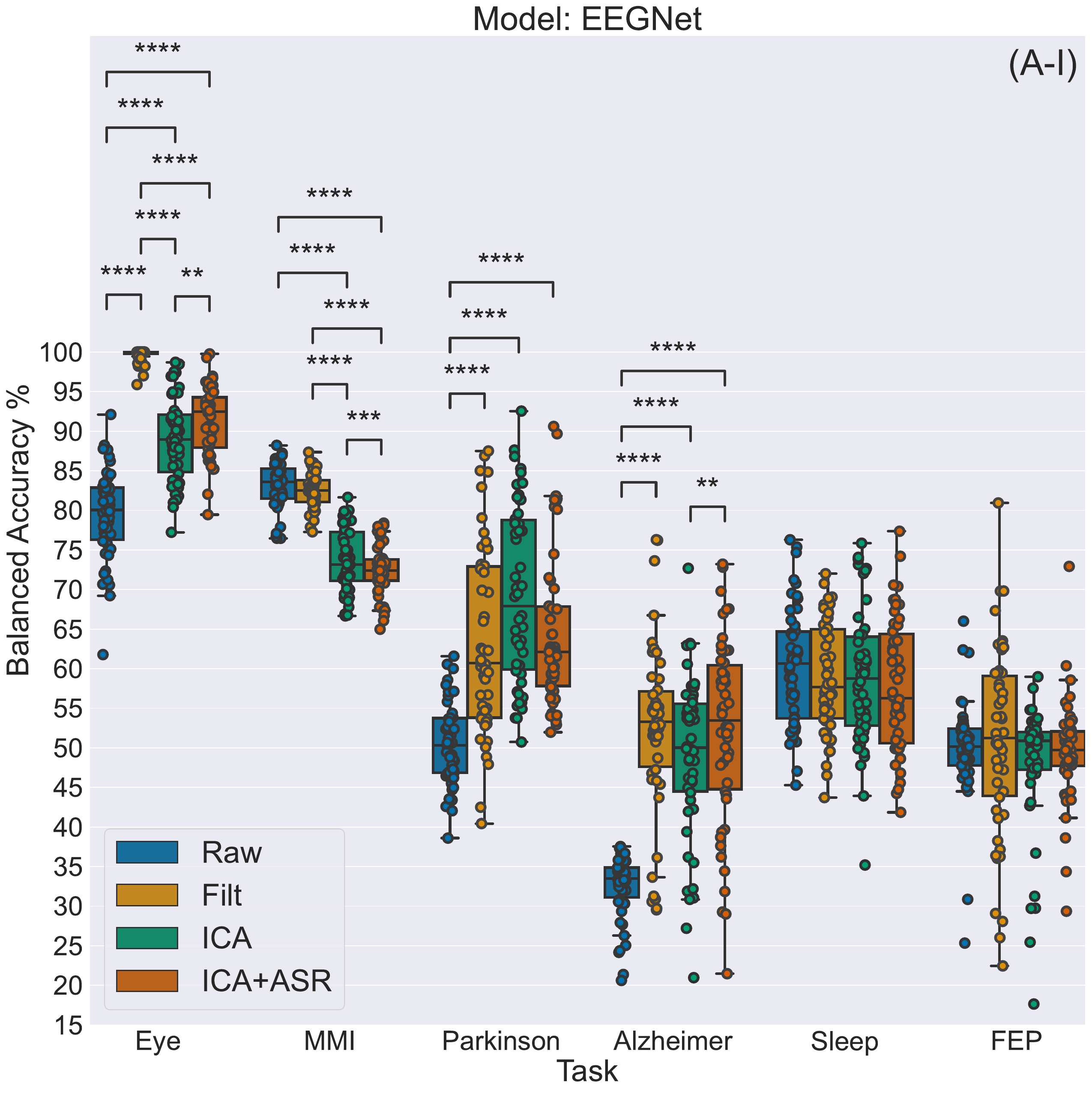}
    \includegraphics[width=0.47\textwidth]{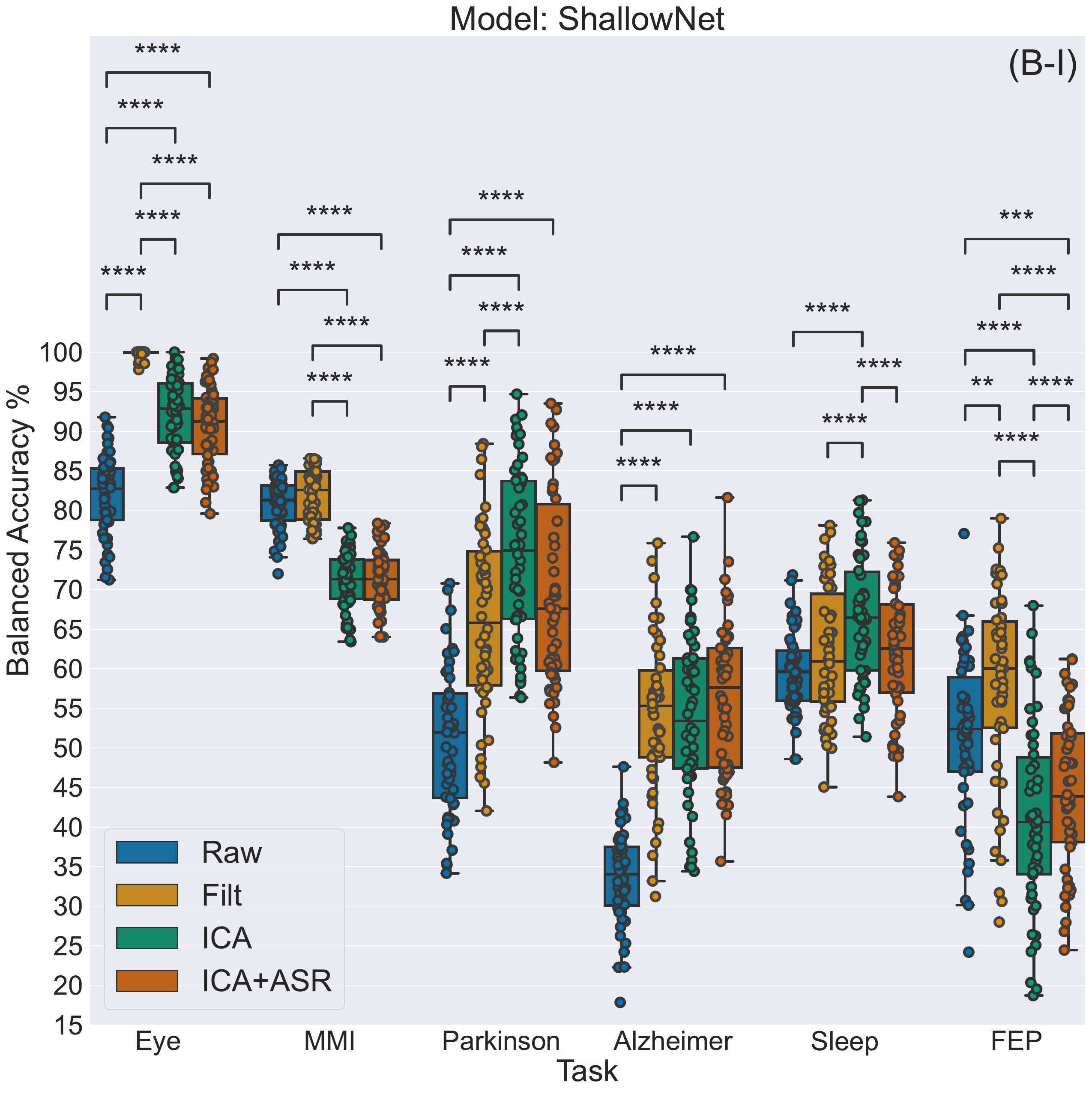}
    
    \includegraphics[width=0.47\textwidth]{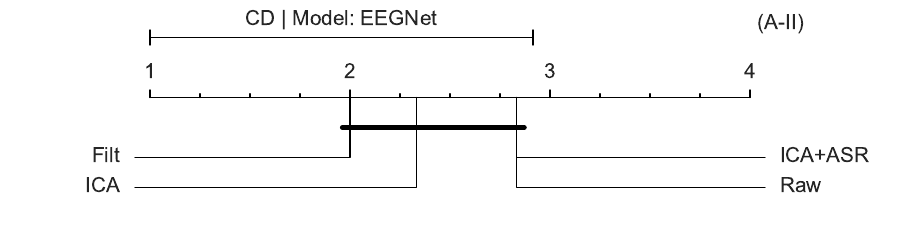}
    \includegraphics[width=0.47\textwidth]{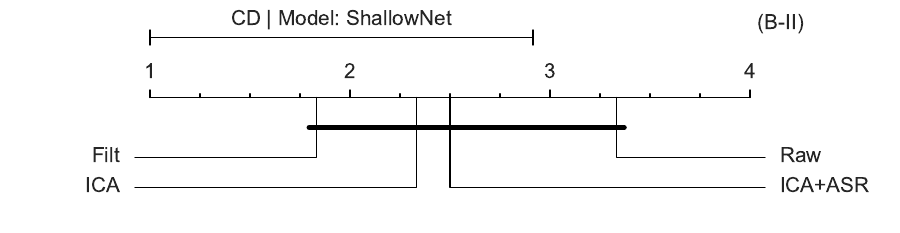}
    
    \includegraphics[width=0.47\textwidth]{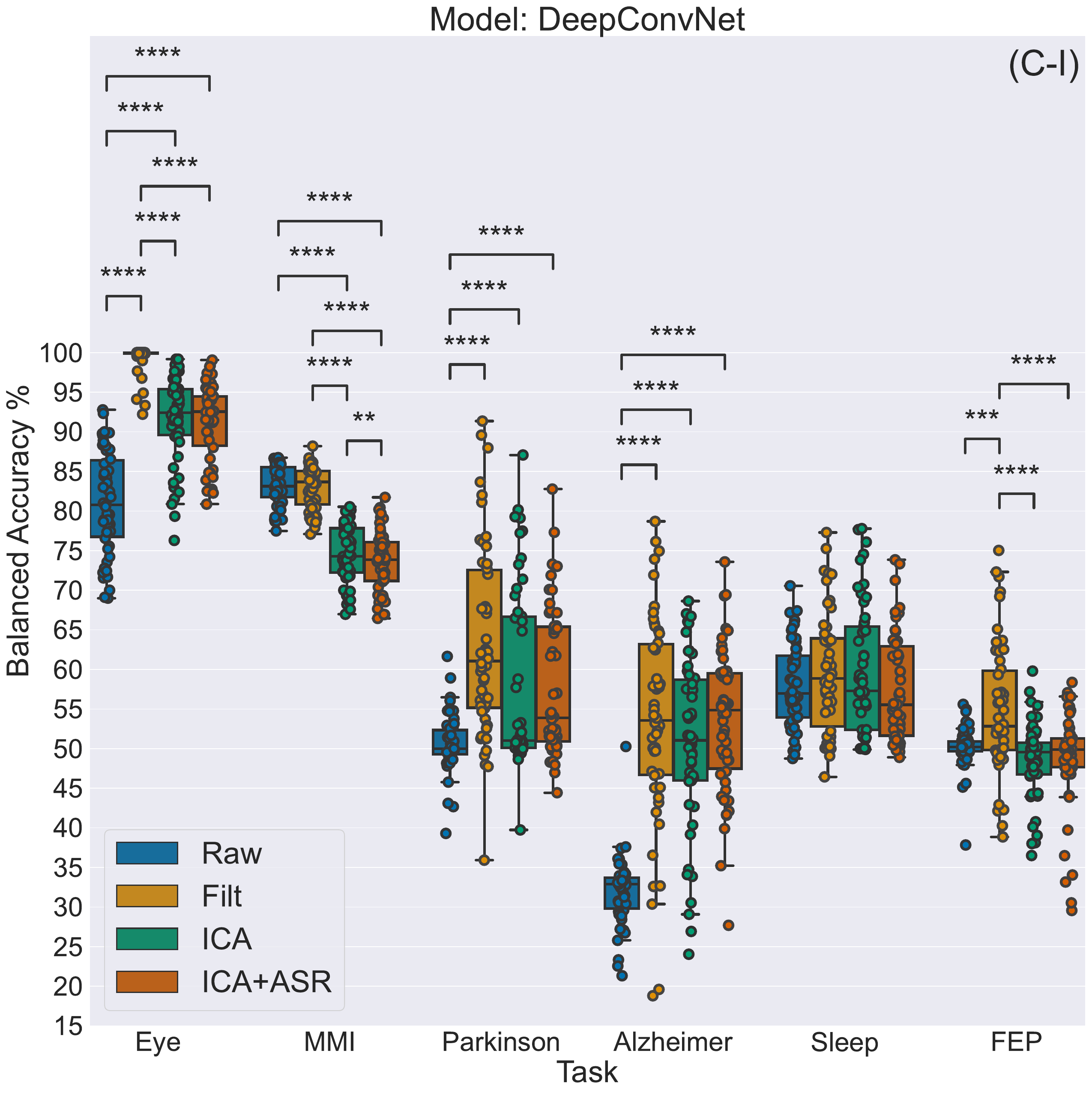}
    \includegraphics[width=0.47\textwidth]{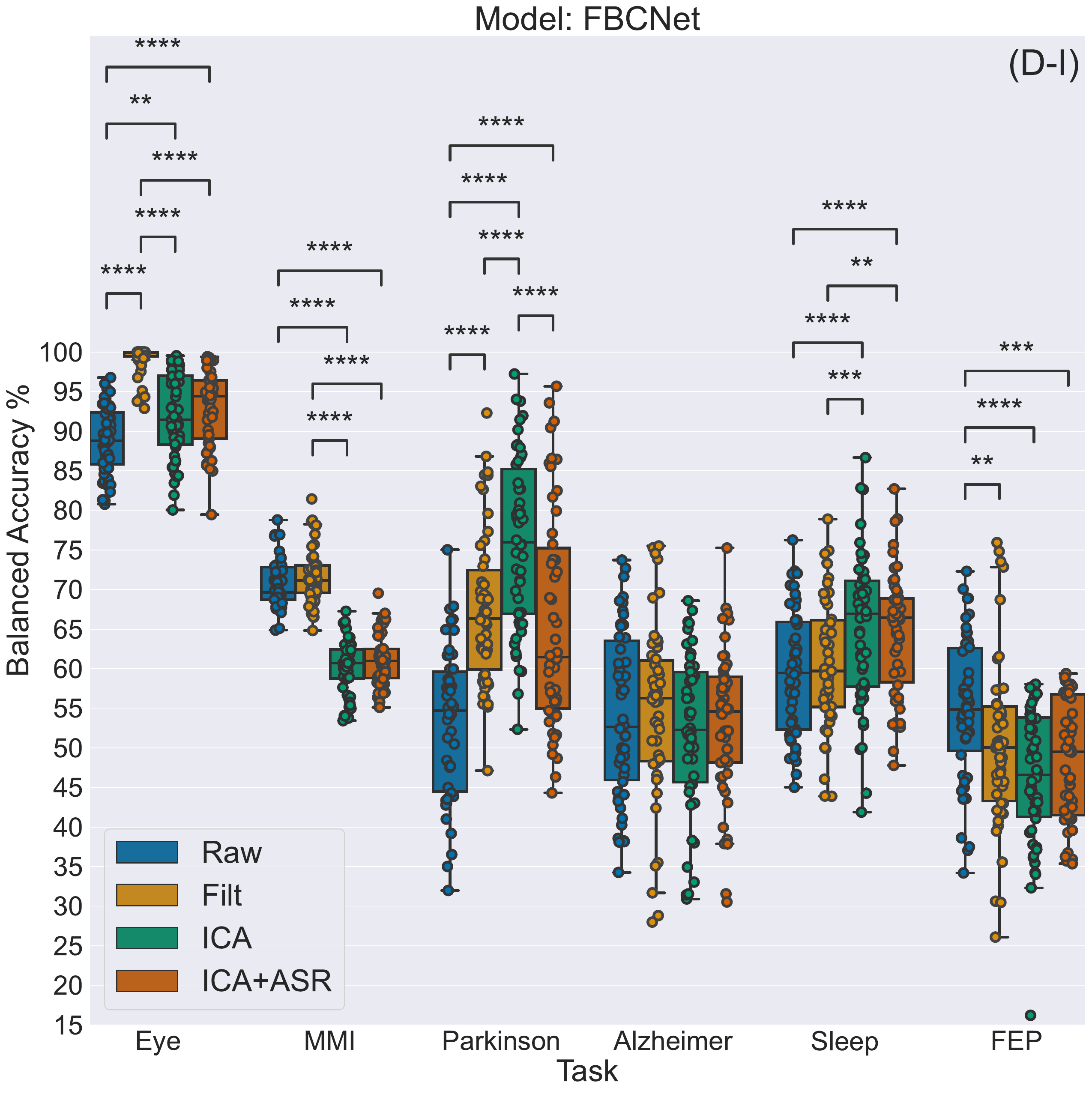}
    
    \includegraphics[width=0.47\textwidth]{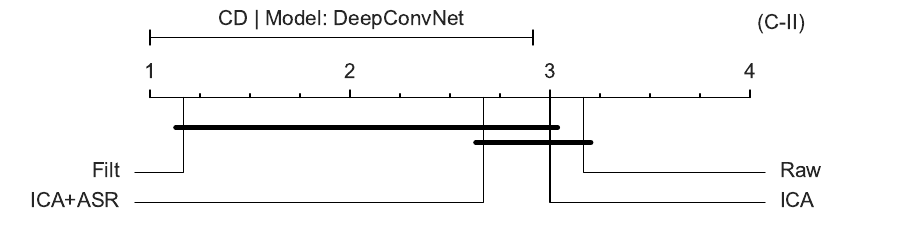}
    \includegraphics[width=0.47\textwidth]{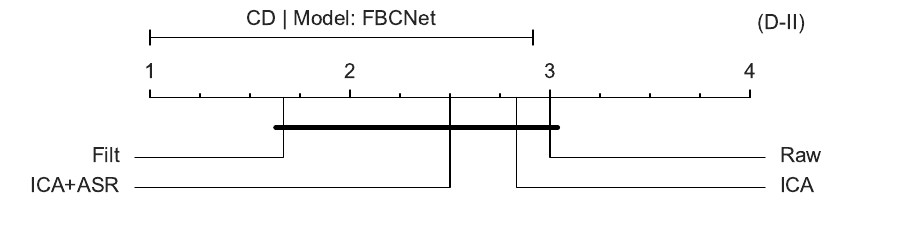}
    \caption{Panels A to D (I) shows the balanced accuracies for the 6 tasks and for the 4 pipelines investigated.
    Results are marginalized by deep learning models which are: in Panel A EEGNet,  in Panel B ShallowNet, in Panel C DeepConvNet and in Panel D FBCnet.
    Asterisks indicate significant results from Wilcoxon's signed rank tests, performed within the same task: ($**$$**$) $p<0.0001$; ($**$$*$) $p<0.001$; ($**$) $p<0.01$ and ($*$) $p<0.05$.
    Results are Holm corrected, for multiple comparison within the same task.
    Panels A to D (II) shows the critical distance (CD) diagram for the 4 pipelines marginalized by four models.
    All classifiers are compared against each other with the Nemenyi's test.
    Groups of pipelines that are not significantly different ($p<0.05$) are connected.}
    \label{fig: models_results}
\end{figure*}

% ----- CI LAVORA ANDREA ------
% intro breve sulla sezione 
Differences were found between the investigated pipelines, both at the intra-task and the inter-task level.
The balanced accuracies of the all 4800 trained models, are reported in \autoref{fig: models_results}.
Results are reported in four panels, one for each deep learning architecture, respectively: A-EEGNet, B-ShallowNet, C-DeepConvNet and in D-FBCNet.
Considering the intra-task level of analysis, only strong significant results ($p<0.001$) from the Wilcoxon tests are reported and discussed in the next paragraphs.

\begin{figure}[!t]
    \centering
    \includegraphics[width=0.8\linewidth]{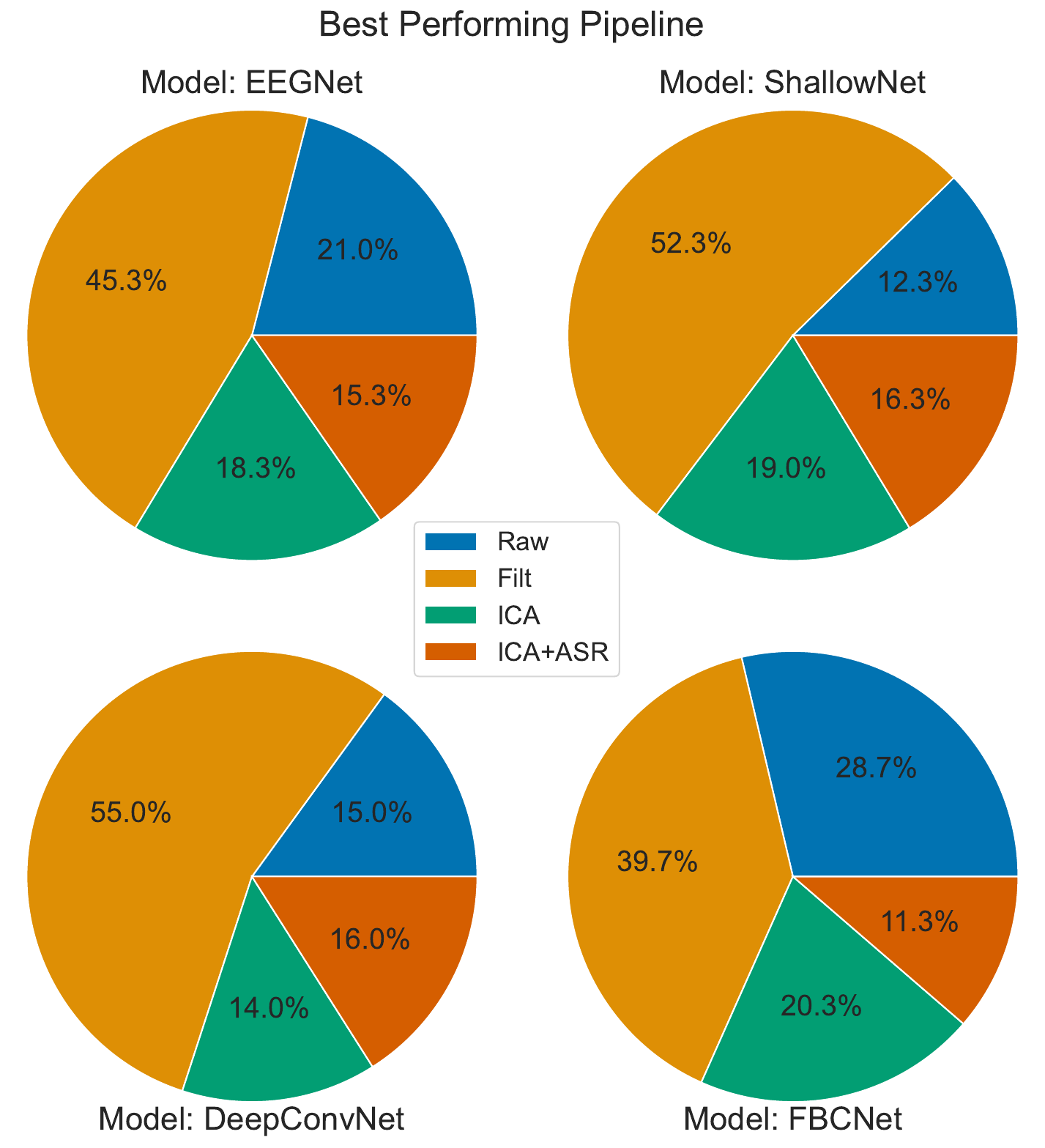}
    \caption{Best performing pipeline; results are marginalized by model. 
    The pie-charts report the number of times (in \%) a pipeline is better than the others, over 300 results (6 tasks $\times$ 50 folds).}
    \label{fig: pie_results}
\end{figure}

\subsection{EEGNet}
% risultati marginalizzati per eegnet
\autoref{fig: models_results}, Panels A I-II, shows the results for EEGNet.
Wilcoxon tests shows that Raw is significantly worse than the other three pipelines in the Eye, Parkinson and Alzheimer task; Raw and Filt are better than the other two in the MMI task; for the Sleep and FEP tasks there are no significant differences between pipelines.
The Friedman' statistic $\chi_F^2=1.8$ gives not statistically significant results ($p=0.61$, under $\chi^2$ assumption).
The average ranks for the four pipelines are: $R_{\text{Raw}}=2.83$, $R_{\text{Filt}}=2$, $R_{\text{ICA}}=2.33$ and $R_{\text{ICA+ASR}}=2.83$, where Filt turns out to be the best performing pipeline.
The maximal difference between average ranks is $R_{\text{Filt-Raw}}=R_{\text{Filt - ICA+ASR}}=0.83$, between Filt and the Raw and ICA+ASR pipelines.

\subsection{ShallowNet}
% risultati marginalizzati per shallownet
\autoref{fig: models_results}, Panels B I-II, shows the results for ShallowNet.
Wilcoxon tests shows that Raw is significantly worse than the other three pipelines in the Eye, Parkinson and Alzheimer task; Raw and Filt are better than the other two in the MMI and in the FEP task; ICA is better than the other three pipelines for the Sleep task.
The Friedman' statistic $\chi_F^2=4.2$ gives not statistically significant results ($p=0.24$, under $\chi^2$ assumption).
The average ranks for the four pipelines are: $R_{\text{Raw}}=3.33$, $R_{\text{Filt}}=1.83$, $R_{\text{ICA}}=2.33$ and $R_{\text{ICA+ASR}}=2.5$, where Filt turns out to be the best performing pipeline.
The maximal difference between average ranks is $R_{\text{Filt-Raw}}=1.5$, between the Filt and the Raw pipelines.

\subsection{DeepConvNet}
% risultati marginalizzati per deepconvnet
\autoref{fig: models_results}, Panels C I-II, shows the results for DeepConvNet.
Wilcoxon tests shows that Raw is significantly worse than the other three pipelines in the Eye, Parkinson and Alzheimer task; Raw and Filt are better than the other two in the MMI and in the FEP task; Filt is better than the other three pipelines for the FEP task; for the Sleep task there are no significant differences between pipelines.
The Friedman' statistic $\chi_F^2=9.0$ gives statistically significant results at $\alpha=0.05$ ($p=0.029$, under $\chi^2$ assumption). 
The average ranks for the four pipelines are: $R_{\text{Raw}}=3.17$, $R_{\text{Filt}}=1.17$, $R_{\text{ICA}}=3.0$ and $R_{\text{ICA+ASR}}=2.67$, where Filt turns out to be the best performing pipeline.
The maximal difference between average ranks is $R_{\text{Filt-Raw}}=2$, between the Filt and the Raw pipelines.

\subsection{FBCNet}
% risultati marginalizzati per fbcnet
\autoref{fig: models_results}, Panels D I-II, shows the results for FBCNet.
Wilcoxon tests shows that Raw is significantly worse than the other three pipelines in the Eye and Parkinson task; Raw and Filt are better than the other two in the MMI task; ICA is better than the other three pipelines for the Parkinson task; ICA is better than Raw and Filt for the Sleep task; Raw is better than ICA and ICA+ASR for the FEP task; for the Alzheimer task there are no significant differences between pipelines.
The Friedman' statistic $\chi_F^2=3.8$ gives not statistically significant results ($p=0.28$, under $\chi^2$ assumption). 
The average ranks for the four pipelines are: $R_{\text{Raw}}=3$, $R_{\text{Filt}}=1.67$, $R_{\text{ICA}}=2.83$ and $R_{\text{ICA+ASR}}=2.5$, where Filt turns out to be the best performing pipeline.
The maximal difference between average ranks is $R_{\text{Filt-Raw}}=1.33$, between the Filt and the Raw pipelines.

\section{Discussion}
\label{sec:discussion}
% ----- CI LAVORA ANDREA, RIFORMULATO DA PUP ------
% ----- POSSIBILE VARIANTE DELLA DISCUSSIONE ------ 

% 1) Intro on the role of preprocessing and the current situation 
%Applying deep learning to time series, especially biomedical signals such as EEG, faces many challenges. 
Researchers must deal with many degrees of freedom when working with EEG data, and preprocessing is undoubtedly a key one.
If in the medical field the motto ``the more, the better" is usually followed, with a large research community devoted to constantly improving the efficacy of automatic pipelines, when it comes to deep learning, things are not so clear-cut.
Thus, the goal of this work was to shed light on the topic and investigate whether preprocessing changes the performance of deep learning models in a relevant way.

%2) Comparison with things said in the introduction and new discoveries from our analysis
While previous research work has demonstrated that DL can work with raw data for certain applications, the presented analysis shows how better results can be achieved if data are at least filtered.
The same applies for a more intense preprocessing.
While the investigation of specific diseases (e.g., Parkinson's, Alzheimer's) can benefit from the addition of automatic artifact handling techniques, the results indicate that a minimal filtering represents the optimal trade-off between computational requirements and achieved accuracy (see also Figure 1 of the supplementary material).

% 3) The importance of certain choices that makes this analysis solid and valid
Concerning the interpretation of the results presented in \ref{sec:results}, it is important to further clarify the relevance of data splitting.
As this work aims to find the effects of data preprocessing on DL performance, extra care is needed when measuring and comparing the performance of deep neural networks. 
A properly designed DL application requires separate training, validation, and independent test sets. 
Improper data splitting can lead to overly optimistic results. 
For instance, splitting at the window level allows EEG windows from the same subject to appear in both training and test sets, introducing bias. 
Future windows may end up in the train set, leading to scenarios where future data predict past events. 
Additionally, consecutive windows, even if separated in time, can remain highly correlated, further skewing results. 
Subject-specific information could be present throughout the recording, making windows easily predictable and overestimating test metrics (see section A.3 of the supplementary material).
To address this, splitting the data by subject helps to achieve a more unbiased estimate of performance, although the way in which subjects are allocated in the different sets still affects the performance of the model. 
N-LNSO allows for repeated measures of test accuracy across different sets, thereby increasing reliability.

When repeated measures are available, t-tests are often used to assess differences between groups. 
However, Dietterich's work~\cite{Dietterich} shows that t-tests on k-results from k-fold cross-validation yield inflated p-values due to underestimated variance.
Nadeau and Bengio, 2005~\cite{BengioInf} introduced a correction factor based on the ratio of samples between training and test sets, but their method assumes a constant ratio and ignores the validation set, which is not applicable in many DL applications. 
In contrast, the Friedman test evaluates relative ranks of classifier performances rather than numerical differences. 
Subject splitting combined with N-LNSO provides an unbiased estimate of median test accuracy, and the Friedman test offers realistic p-values for performance differences.

% 4) What our results say?
% 4.1) Introduction: EEG preprocessing has an impact on performances and cannot be neglected.
%Concerning the results obtained, it is clear that the EEG preprocessing has an impact on model performance and this degree of freedom cannot be neglected by DL researchers.
Concerning the results obtained, it is clear that the EEG preprocessing significantly impacts model performance, and this degree of freedom must be considered by DL researchers.

% 4.2) Task Level: (Wilcoxon results).
For example, when performing the Parkinson's task with EEGNet (\autoref{fig: models_results} Panel A-I, task Parkinson), the median value using raw data is near 50\%, while using the ICA pipeline the value is around 67\%.
Again, when performing the Eye task with EEGNet (\autoref{fig: models_results} Panel A-I, task Eye), the median value using raw data is near 80\%, while using the Filt pipeline it reaches almost 100\%.
Looking at the Wilcoxon signed rank test, significant results are found independently from the model considered, confirming what was said.
However, the goal of this paper is to go beyond pair-wise comparisons.
Considering an architecture-specific point of view, Friedman's test does not always give significant results. 
\autoref{fig: pie_results} clearly demonstrates that the Filt pipeline gives the best results in more than half of the splits over the six tasks.
The Critical Difference diagrams in \autoref{fig: models_results} Panels A-D (II), further confirm this, where Filt is always ranked first, above the other three pipelines.
At the same time, Friedman's test gives significant result only for DeepConvNet.
% 4.4) In general: depends on the overall final results
% Consequently, while Filt significantly outperforms the Raw pipeline, it remains unclear whether this improvement stems from the pipeline's inherent superiority, from the architecture's sensitivity to EEG preprocessing or from other reasons.
Consequently, while Filt significantly outperforms the Raw pipeline, it remains unclear whether this improvement is due to the inherent superiority of the pipeline, the sensitivity of the architecture to EEG preprocessing, or other reasons.

%5) Future Works
This question can be further investigated in future works with some insights from this paper.
First, based on the performances of the Filt pipeline, it is plausible that increasing the number of tasks, and thus diminishing the critical difference (CD), other architectures will show significance on Friedman's test.
Consequently, it would confirm the hypothesis that the pipeline is better, independently from the task.
This would put the problem back to the low-power of Friedman's test.

At the same time, it would be interesting to understand the spectral properties of DL-architectures. 
This could be done by convolving specific input signals with the trained kernels.
While this is true in general, it is especially useful for pathology classification tasks, such as Alzheimer's and Parkinson's, where from a medical point of view, lot of attention is put to the Power Spectral Density (PSD) and the ratio between power bands.
For instance, Alzheimer's disease is associated with increased power in the theta band following alpha rhythm desynchronization.
This analysis could explain why DeepConvNet, is sensitive to preprocessing, which is known to enhance the PSD of the EEG recording by increasing its signal-to-noise ratio.

%6) Conclusions
%In conclusion, deep neural networks are powerful models with incredible flexibility and ability to achieve top performance in various tasks.
%However, this potential should not be misunderstood as the network's ability to solve any problem without having to worry about managing it.

\section{Conclusion}
\label{sec:conclusion}
% ----- CI LAVORA PUP ------
In this work, the role of preprocessing in EEG-DL applications was investigated.
Four preprocessing pipelines with different levels of complexity were considered, ranging from raw data to a richer preprocessing with established artifact handling automated algorithms (e.g., ICLabel, ASR).
The analysis was performed on six representative tasks that cover a wide range of possible clinical and non-clinical use cases, using four architectures among the most often used in EEG.
%Performance metrics from 4800 trainings were gathered with an unbiased cross validation strategy and used to look for statistical differences both at the local (task) and the model (multiple tasks, model marginalization) level.
%Significant results were found at the task level independently from the model considered; differences were also found globally on the largest investigated model, DeepConvNet.
Performance metrics from 4800 trainings were gathered with an unbiased cross validation strategy and used to look for statistical differences both at the intra-task and the inter-task level.
Significant results were found at the intra-task level independently from the model considered; inter-task differences were also found on the largest investigated model, DeepConvNet.

Furthermore, the analysis highlighted an overall trend characterized by raw data generally bringing to underperforming models (always getting last in terms of average ranking) and a minimal preprocessing without artifact handling being superior, always ranking first on average.
Models do not always benefit from richer pipelines, suggesting a potential discriminatory power of EEG artifacts and leaving the question of which properties the DL-architecture is learning still unsolved.

%appendices
%Appendixes, if needed, appear before the acknowledgment.

\section{Acknowledgment}
\textit{Contributions}: 
FDP - Conceptualization, Data Preprocessing (ds004362), Model Training, Writing - Original Draft. AZ - Conceptualization, Data Preprocessing, Statistical Analysis, Writing - Original Draft. MA - Project Administration, Supervision. All authors reviewed and edited previous versions of the manuscript.

\textit{Code and data availability}: code used to produce both results and figures is available at \url{https://github.com/MedMaxLab/eegprepro}. All data that support the findings of this study are openly available within the OpenNeuro platform.

The authors have no competing interests to declare that are relevant to the content of this article.

%\section{References}
\bibliographystyle{ieeeRBE}
\bibliography{bibliography}

\begin{thebibliography}{10}
\providecommand{\url}[1]{#1}
\csname url@samestyle\endcsname
\providecommand{\newblock}{\relax}
\providecommand{\bibinfo}[2]{#2}
\providecommand{\BIBentrySTDinterwordspacing}{\spaceskip=0pt\relax}
\providecommand{\BIBentryALTinterwordstretchfactor}{4}
\providecommand{\BIBentryALTinterwordspacing}{\spaceskip=\fontdimen2\font plus
\BIBentryALTinterwordstretchfactor\fontdimen3\font minus \fontdimen4\font\relax}
\providecommand{\BIBforeignlanguage}[2]{{%
\expandafter\ifx\csname l@#1\endcsname\relax
\typeout{** WARNING: IEEEtran.bst: No hyphenation pattern has been}%
\typeout{** loaded for the language `#1'. Using the pattern for}%
\typeout{** the default language instead.}%
\else
\language=\csname l@#1\endcsname
\fi
#2}}
\providecommand{\BIBdecl}{\relax}
\BIBdecl

\bibitem{BCIRew}
R.~Abiri, S.~Borhani, E.~W. Sellers, Y.~Jiang, and X.~Zhao, ``A comprehensive review of {EEG}-based brain-computer interface paradigms,'' \emph{Journal of neural engineering}, vol.~16, no.~1, p. 011001, 2019.

\bibitem{EmotionRew}
X.~Li \emph{et~al.}, ``{EEG} based emotion recognition: A tutorial and review,'' \emph{ACM Computing Surveys}, vol.~55, no.~4, pp. 1--57, 2022.

\bibitem{SleepRw}
K.~A.~I. Aboalayon, M.~Faezipour, W.~S. Almuhammadi, and S.~Moslehpour, ``Sleep stage classification using {EEG} signal analysis: a comprehensive survey and new investigation,'' \emph{Entropy}, vol.~18, no.~9, p. 272, 2016.

\bibitem{EpilepsyRew}
K.~Rasheed \emph{et~al.}, ``Machine learning for predicting epileptic seizures using {EEG} signals: A review,'' \emph{IEEE reviews in biomedical engineering}, vol.~14, pp. 139--155, 2020.

\bibitem{ParkinsonRew}
A.~M. Maitin, J.~P. Romero~Mu{\~n}oz, and {\'A}.~J. Garc{\'\i}a-Tejedor, ``Survey of machine learning techniques in the analysis of {EEG signals for Parkinson’s disease}: A systematic review,'' \emph{Applied Sciences}, vol.~12, no.~14, p. 6967, 2022.

\bibitem{AlzheimerRew}
Y.~Zhao and L.~He, ``Deep learning in the {EEG diagnosis of Alzheimer’s disease},'' in \emph{Computer Vision-ACCV 2014 Workshops: Singapore, Singapore, November 1-2, 2014, Revised Selected Papers, Part I 12}.\hskip 1em plus 0.5em minus 0.4em\relax Springer, 2015, pp. 340--353.

\bibitem{DLReviewEEG}
A.~Craik, Y.~He, and J.~L. Contreras-Vidal, ``Deep learning for electroencephalogram {(EEG)} classification tasks: a review,'' \emph{Journal of neural engineering}, vol.~16, no.~3, p. 031001, 2019.

\bibitem{DLTrio}
Y.~LeCun, Y.~Bengio, and G.~Hinton, ``Deep learning,'' \emph{nature}, vol. 521, no. 7553, pp. 436--444, 2015.

\bibitem{EEGDataRew}
C.~P. Da~Silva, S.~Tedesco, and B.~O’Flynn, ``{EEG} datasets for healthcare: a scoping review,'' \emph{IEEE Access}, 2024.

\bibitem{happe}
L.~J. Gabard-Durnam, A.~S. Mendez~Leal, C.~L. Wilkinson, and A.~R. Levin, ``The harvard automated processing pipeline for electroencephalography {(HAPPE)}: standardized processing software for developmental and high-artifact data,'' \emph{Frontiers in neuroscience}, vol.~12, p. 316496, 2018.

\bibitem{automagic}
A.~Pedroni, A.~Bahreini, and N.~Langer, ``Automagic: Standardized preprocessing of big {EEG} data,'' \emph{NeuroImage}, vol. 200, pp. 460--473, 2019.

\bibitem{BIDSalign}
(accepted), A.~Zanola, F.~Del~Pup, C.~Porcaro, and M.~Atzori, ``{BIDSAlign: a library for automatic merging and preprocessing of multiple EEG repositories},'' \emph{Journal of Neural Engineering}, 2024.

\bibitem{ssleeg}
M.~H. Rafiei, L.~V. Gauthier, H.~Adeli, and D.~Takabi, ``Self-supervised learning for electroencephalography,'' \emph{IEEE Transactions on Neural Networks and Learning Systems}, 2022.

\bibitem{Yannick}
\BIBentryALTinterwordspacing
Y.~Roy, H.~Banville, I.~Albuquerque, A.~Gramfort, T.~H. Falk, and J.~Faubert, ``Deep learning-based electroencephalography analysis: a systematic review,'' \emph{Journal of Neural Engineering}, vol.~16, no.~5, p. 051001, aug 2019. [Online]. Available: \url{https://doi.org/10.1088/1741-2552/ab260c}
\BIBentrySTDinterwordspacing

\bibitem{Aznan}
N.~K.~N. Aznan, S.~Bonner, J.~Connolly, N.~Al~Moubayed, and T.~Breckon, ``On the classification of {SSVEP}-based dry-{EEG} signals via convolutional neural networks,'' in \emph{2018 IEEE international conference on systems, man, and cybernetics (SMC)}.\hskip 1em plus 0.5em minus 0.4em\relax IEEE, 2018, pp. 3726--3731.

\bibitem{Almogbel}
M.~A. Almogbel, A.~H. Dang, and W.~Kameyama, ``Eeg-signals based cognitive workload detection of vehicle driver using deep learning,'' in \emph{2018 20th International Conference on Advanced Communication Technology (ICACT)}.\hskip 1em plus 0.5em minus 0.4em\relax IEEE, 2018, pp. 256--259.

\bibitem{EEGInterp}
J.~Cui, L.~Yuan, Z.~Wang, R.~Li, and T.~Jiang, ``Towards best practice of interpreting deep learning models for {EEG-based} brain computer interfaces,'' \emph{Frontiers in Computational Neuroscience}, vol.~17, 2023.

\bibitem{Hefron}
R.~Hefron, B.~Borghetti, C.~Schubert~Kabban, J.~Christensen, and J.~Estepp, ``Cross-participant {EEG-based} assessment of cognitive workload using multi-path convolutional recurrent neural networks,'' \emph{Sensors}, vol.~18, no.~5, p. 1339, 2018.

\bibitem{PreproWL}
K.~Kingphai and Y.~Moshfeghi, ``On {EEG} preprocessing role in deep learning effectiveness for mental workload classification,'' in \emph{Human Mental Workload: Models and Applications: 5th International Symposium, H-WORKLOAD 2021, Virtual Event, November 24--26, 2021, Proceedings 5}.\hskip 1em plus 0.5em minus 0.4em\relax Springer, 2021, pp. 81--98.

\bibitem{OpenNeuro}
C.~J. Markiewicz \emph{et~al.}, ``The {OpenNeuro} resource for sharing of neuroscience data,'' \emph{eLife}, vol.~10, p. e71774, oct 2021.

\bibitem{eeguide}
\BIBentryALTinterwordspacing
S.~R. Sinha \emph{et~al.}, ``American clinical neurophysiology society guideline 1: minimum technical requirements for performing clinical electroencephalography,'' \emph{Journal of Clinical Neurophysiology}, vol.~33, no.~4, pp. 303--307, 2016. [Online]. Available: \url{https://doi.org/10.1097/WNP.0000000000000308}
\BIBentrySTDinterwordspacing

\bibitem{MMIRew}
H.~Altaheri \emph{et~al.}, ``Deep learning techniques for classification of electroencephalogram ({EEG}) motor imagery ({MI}) signals: A review,'' \emph{Neural Computing and Applications}, vol.~35, no.~20, pp. 14\,681--14\,722, 2023.

\bibitem{ds004148}
\BIBentryALTinterwordspacing
Y.~Wang, W.~Duan, D.~Dong, L.~Ding, and X.~Lei, ``A test-retest resting and cognitive state {EEG} dataset,'' 2022. [Online]. Available: \url{https://doi.org/10.18112/openneuro.ds004148.v1.0.1}
\BIBentrySTDinterwordspacing

\bibitem{ds004504}
\BIBentryALTinterwordspacing
A.~Miltiadous \emph{et~al.}, ``A dataset of {EEG} recordings from: {Alzheimer's disease}, frontotemporal dementia and healthy subjects,'' 2023. [Online]. Available: \url{https://doi.org/10.18112/openneuro.ds004504.v1.0.6}
\BIBentrySTDinterwordspacing

\bibitem{mmi_rm1}
C.~M. Köllőd, A.~Adolf, K.~Iván, G.~Márton, and I.~Ulbert, ``Deep comparisons of neural networks from the {EEGNet} family,'' \emph{Electronics}, vol.~12, no.~12, 2023.

\bibitem{mmi_rm2}
G.~Zoumpourlis and I.~Patras, ``Motor imagery decoding using ensemble curriculum learning and collaborative training,'' 2024.

\bibitem{ds002778}
\BIBentryALTinterwordspacing
A.~P. Rockhill, N.~Jackson, J.~George, A.~Aron, and N.~C. Swann, ``{UC San Diego resting state EEG} data from patients with {Parkinson's} disease,'' 2021. [Online]. Available: \url{https://doi.org/10.18112/openneuro.ds002778.v1.0.5}
\BIBentrySTDinterwordspacing

\bibitem{ds003490}
\BIBentryALTinterwordspacing
J.~F. Cavanagh, ``{EEG}: 3-stim auditory oddball and rest in {Parkinson's},'' 2021. [Online]. Available: \url{https://doi.org/10.18112/openneuro.ds003490.v1.1.0}
\BIBentrySTDinterwordspacing

\bibitem{ds004902}
\BIBentryALTinterwordspacing
C.~Xiang, X.~Fan, D.~Bai, K.~Lv, and X.~Lei, ``A resting-state {EEG} dataset for sleep deprivation,'' 2024. [Online]. Available: \url{https://doi.org/10.18112/openneuro.ds004902.v1.0.5}
\BIBentrySTDinterwordspacing

\bibitem{ds003947}
\BIBentryALTinterwordspacing
D.~Salisbury, D.~Seebold, and B.~Coffman, ``{EEG}: First episode psychosis vs. control resting task 2,'' 2022. [Online]. Available: \url{https://doi.org/10.18112/openneuro.ds003947.v1.0.1}
\BIBentrySTDinterwordspacing

\bibitem{eeglab}
A.~Delorme and S.~Makeig, ``{EEGLAB}: an open source toolbox for analysis of single-trial {EEG} dynamics including independent component analysis,'' \emph{Journal of Neuroscience Methods}, vol. 134, no.~1, pp. 9--21, 2004.

\bibitem{infomaxica}
A.~J. Bell and T.~J. Sejnowski, ``An information-maximization approach to blind separation and blind deconvolution,'' \emph{Neural Computation}, vol.~7, no.~6, pp. 1129--1159, 11 1995.

\bibitem{ICLabel}
L.~Pion-Tonachini, K.~Kreutz-Delgado, and S.~Makeig, ``{ICLabel}: An automated electroencephalographic independent component classifier, dataset, and website,'' \emph{NeuroImage}, vol. 198, pp. 181--197, 2019.

\bibitem{asr}
T.~R. Mullen \emph{et~al.}, ``Real-time neuroimaging and cognitive monitoring using wearable dry {EEG},'' \emph{IEEE Transactions on Biomedical Engineering}, vol.~62, no.~11, pp. 2553--2567, 2015.

\bibitem{Spherical}
S.~S. Kang, T.~J. Lano, and S.~R. Sponheim, ``Distortions in {EEG} interregional phase synchrony by spherical spline interpolation: causes and remedies,'' \emph{Neuropsychiatric Electrophysiology}, vol.~1, pp. 1--17, 2015.

\bibitem{datanorm}
A.~Apicella, F.~Isgr{\`o}, A.~Pollastro, and R.~Prevete, ``On the effects of data normalization for domain adaptation on {EEG} data,'' \emph{Engineering Applications of Artificial Intelligence}, vol. 123, p. 106205, 2023.

\bibitem{Selfeeg}
F.~Del~Pup, A.~Zanola, L.~F. Tshimanga, P.~E. Mazzon, and M.~Atzori, ``{SelfEEG}: A {Python} library for self-supervised learning in electroencephalography,'' \emph{Journal of Open Source Software}, vol.~9, no.~95, p. 6224, Mar. 2024.

\bibitem{pytorch}
A.~Paszke \emph{et~al.}, ``Pytorch: An imperative style, high-performance deep learning library,'' \emph{Advances in neural information processing systems}, vol.~32, 2019.

\bibitem{scipy}
P.~Virtanen \emph{et~al.}, ``{SciPy} 1.0: fundamental algorithms for scientific computing in python,'' \emph{Nature methods}, vol.~17, no.~3, pp. 261--272, 2020.

\bibitem{seaborn}
M.~L. Waskom, ``Seaborn: statistical data visualization,'' \emph{Journal of Open Source Software}, vol.~6, no.~60, p. 3021, 2021.

\bibitem{statann}
\BIBentryALTinterwordspacing
F.~Charlier \emph{et~al.}, ``Statannotations,'' Oct. 2022. [Online]. Available: \url{https://doi.org/10.5281/zenodo.7213391}
\BIBentrySTDinterwordspacing

\bibitem{eegnet}
V.~J. Lawhern, A.~J. Solon, N.~R. Waytowich, S.~M. Gordon, C.~P. Hung, and B.~J. Lance, ``{EEGNet}: a compact convolutional neural network for {EEG}-based brain--computer interfaces,'' \emph{Journal of neural engineering}, vol.~15, no.~5, p. 056013, 2018.

\bibitem{shallow}
R.~T. Schirrmeister \emph{et~al.}, ``Deep learning with convolutional neural networks for {EEG} decoding and visualization,'' \emph{Human brain mapping}, vol.~38, no.~11, pp. 5391--5420, 2017.

\bibitem{fbcnet}
R.~Mane, N.~Robinson, A.~P. Vinod, S.-W. Lee, and C.~Guan, ``A multi-view {CNN} with novel variance layer for motor imagery brain computer interface,'' in \emph{2020 42nd Annual International Conference of the IEEE Engineering in Medicine \& Biology Society (EMBC)}, 2020, pp. 2950--2953.

\bibitem{SubjAwSSL}
J.~Y. Cheng, H.~Goh, K.~Dogrusoz, O.~Tuzel, and E.~Azemi, ``Subject-aware contrastive learning for biosignals,'' \emph{arXiv preprint arXiv:2007.04871}, 2020.

\bibitem{LOSO}
S.~Kunjan \emph{et~al.}, ``The necessity of leave one subject out {(LOSO)} cross validation for {EEG} disease diagnosis,'' in \emph{Brain Informatics: 14th International Conference, BI 2021, Virtual Event, September 17--19, 2021, Proceedings 14}.\hskip 1em plus 0.5em minus 0.4em\relax Springer, 2021, pp. 558--567.

\bibitem{overfitting}
G.~C. Cawley and N.~L. Talbot, ``On over-fitting in model selection and subsequent selection bias in performance evaluation,'' \emph{The Journal of Machine Learning Research}, vol.~11, pp. 2079--2107, 2010.

\bibitem{adam}
D.~Kingma and J.~Ba, ``Adam: A method for stochastic optimization,'' in \emph{International Conference on Learning Representations (ICLR)}, San Diega, CA, USA, 2015.

\bibitem{wilcoxon}
\BIBentryALTinterwordspacing
F.~Wilcoxon, ``Individual comparisons by ranking methods,'' \emph{Biometrics Bulletin}, vol.~1, no.~6, pp. 80--83, 1945. [Online]. Available: \url{http://www.jstor.org/stable/3001968}
\BIBentrySTDinterwordspacing

\bibitem{mannwhit}
H.~B. Mann and D.~R. Whitney, ``{On a Test of Whether one of Two Random Variables is Stochastically Larger than the Other},'' \emph{The Annals of Mathematical Statistics}, vol.~18, no.~1, pp. 50 -- 60, 1947.

\bibitem{holm}
S.~Holm, ``A simple sequentially rejective multiple test procedure,'' \emph{Scandinavian Journal of Statistics}, vol.~6, no.~2, pp. 65--70, 1979.

\bibitem{Friedman}
\BIBentryALTinterwordspacing
M.~Friedman, ``The use of ranks to avoid the assumption of normality implicit in the analysis of variance,'' \emph{Journal of the American Statistical Association}, vol.~32, no. 200, pp. 675--701, 1937. [Online]. Available: \url{https://doi.org/10.1080/01621459.1937.10503522}
\BIBentrySTDinterwordspacing

\bibitem{demsar}
\BIBentryALTinterwordspacing
J.~Dem{\v{s}}ar, ``Statistical comparisons of classifiers over multiple data sets,'' \emph{Journal of Machine Learning Research}, vol.~7, no.~1, pp. 1--30, 2006. [Online]. Available: \url{http://jmlr.org/papers/v7/demsar06a.html}
\BIBentrySTDinterwordspacing

\bibitem{raschka2020model}
S.~Raschka, ``Model evaluation, model selection, and algorithm selection in machine learning,'' 2020.

\bibitem{effectprep}
\BIBentryALTinterwordspacing
E.~Alshdaifat, D.~Alshdaifat, A.~Alsarhan, F.~Hussein, and S.~M. F.~S. El-Salhi, ``The effect of preprocessing techniques, applied to numeric features, on classification algorithms’ performance,'' \emph{Data}, vol.~6, no.~2, 2021. [Online]. Available: \url{https://doi.org/10.3390/data6020011}
\BIBentrySTDinterwordspacing

\bibitem{nemenyi}
P.~B. Nemenyi, \emph{Distribution-free multiple comparisons.}\hskip 1em plus 0.5em minus 0.4em\relax Princeton University, 1963.

\bibitem{Dietterich}
\BIBentryALTinterwordspacing
T.~G. Dietterich, ``Approximate statistical tests for comparing supervised classification learning algorithms,'' \emph{Neural Computation}, vol.~10, no.~7, pp. 1895--1923, 10 1998. [Online]. Available: \url{https://doi.org/10.1162/089976698300017197}
\BIBentrySTDinterwordspacing

\bibitem{BengioInf}
C.~Nadeau and Y.~Bengio, ``Inference for the generalization error,'' \emph{Machine learning}, vol.~52, no.~3, pp. 239--281, 2003.

\end{thebibliography}

\section{Supplementary Material}

\subsection{Additional implementation choices}
This section extends the list of implementation choices.
It covers aspects that are interesting from an engineering point of view, and expands on others already covered in the main body.

\subsubsection{Reducing the training time}
Code optimization plays an important role in the presented analysis.
Considering the large number of trainings - nearly 8000, including those performed to assess the learning rate - saving even a single second for each training easily adds up to hours in total.
Since no new models were implemented for this work, two key factors can influence the time per epoch, and consequently the total one: 1) how fast data are loaded; 2) how fast mini-batches are generated.
Concerning the first point, \textit{BIDSAlign}, as a MATLAB\textsuperscript{\textregistered} library, stores the preprocessed files in a \textit{.mat} format, which is slower to load compared to other format commonly used in a Python environment.
A simple conversion of preprocessed EEG records in \textit{.pickle} format can lead to a huge time gain even if dataloaders do not operate lazily, i.e., do not preload the entire dataset.

Concerning the second point, the optimization procedure basically summarizes in how a sample is extracted from the repository when needed by the dataloader.
Three different scenario may arise.
The first consists in the dataloader loading the entire EEG record each time it needs one of its partitions, which is obviously inefficient.
The second, which is the one proposed in \textit{SelfEEG}, is characterized by a good compromise between computational efficiency and memory allocation. In short, each time an EEG is loaded, it is kept in memory until a new one is needed. When coupled with the custom sampler provided by the same library, an optimized procedure can be employed to minimize the number of file loading operations while maintaining batch heterogeneity.
The third one, also included in \textit{SelfEEG}, simply consists in pre-loading the entire dataset as a huge tensor (or a tuple of tensors), making it possible to load each file only once.

The third scenario is certainly the most efficient but requires a greater amount of RAM memory, even more than the dataset size if overlapping between windows is applied.
If the preloaded dataset is also entirely moved to the GPU, all the mini-batches device conversions (the classical Pytorch's ``.to(device=`cuda')" call) can be skipped.

Overall, it can be concluded that a simple format conversion and a data preloading procedure (whenever possible) can drastically reduce the total run time and allow the execution of a huge number of training in a short amount of time, as required for this experiment.
\autoref{tab: runtime} reports the total and epoch runtime of six different scenario applied to a 100 epoch training of ShallowNet for the Alzheimer's classification task, which is the one with the largest number of samples. It is easy to see how important this type of optimization steps are.
 
\renewcommand{\arraystretch}{1.5} % Default value: 1
\begin{table}[!t]
    \centering
    \caption{Training time with different data loading strategies}
    \begin{tabular}{ccccc}
        \Xhline{2\arrayrulewidth} \\[-1,3em]
        \# & \makecell{Dataloader \\ type} & \makecell{epoch\\ time \\ \text{[s]}} & \makecell{training \\ time \\ \text{[s]}} & \makecell{relative epoch \\ time gain from \\ previous pipeline } \\
        \\[-1,3em] \Xhline{2\arrayrulewidth}
        \multicolumn{1}{c|}{1} & \multicolumn{1}{c|}{\makecell{ \\[-0,6em] basic dataloader \\ no preload \\ .mat files \\[-0,6em] $ $}} & \multicolumn{1}{c|}{5391.89} & \multicolumn{1}{c|}{538499$^*$} & slowest \\
        \hline
        \multicolumn{1}{c|}{2} & \multicolumn{1}{c|}{\makecell{ \\[-0,6em] basic dataloader \\ no preload \\ .pickle files \\[-0,6em] $ $}} & \multicolumn{1}{c|}{1463.94} & \multicolumn{1}{c|}{146584$^*$} & -72.8\% \\
        \hline
        \multicolumn{1}{c|}{3} & \multicolumn{1}{c|}{\makecell{\\[-0,6em] \textit{SelfEEG} dataloader \\ no preload \\ .mat files \\[-0,6em] $ $}} & \multicolumn{1}{c|}{30.47} & \multicolumn{1}{c|}{3083} & -97.9\% \\
        \hline
        \multicolumn{1}{c|}{4} & \multicolumn{1}{c|}{\makecell{ \\[-0,6em] \textit{SelfEEG}  dataloader \\ no preload \\ .pickle files \\[-0,6em] $ $}} & \multicolumn{1}{c|}{11.57} & \multicolumn{1}{c|}{1171} & -62.0\% \\
        \hline
        \multicolumn{1}{c|}{5} & \multicolumn{1}{c|}{\makecell{\\[-0,6em] \textit{SelfEEG}  dataloader \\ preload on CPU \\ .pickle files \\[-0,6em] $ $}} & \multicolumn{1}{c|}{4.58} & \multicolumn{1}{c|}{484} & -60.4\% \\
        \hline
        \multicolumn{1}{c|}{6} & \multicolumn{1}{c|}{\makecell{\\[-0,6em] \textit{SelfEEG}  dataloader \\ preload on GPU \\ .pickle files \\[-0,6em] $ $}} & \multicolumn{1}{c|}{3.58} & \multicolumn{1}{c|}{384} & -21.8\% \\
        \hline
        \multicolumn{5}{l}{${ }^{\text{*estimated from a run with less epochs}}$}
    \end{tabular}
    \label{tab: runtime}
\end{table}
\renewcommand{\arraystretch}{1} % Default value: 1

\subsubsection{Nested Leave-N-Subject-Out split summary}
The Nested Leave-N-Subject-Out (N-LNSO) cross validation was proposed with the specific purpose of getting an unbiased estimate of the centrality measure used for the statistical analysis.
However, given that EEG records of different subjects may also have different length, one might ask if sample and class ratios are preserved between the partitions or, formulating in another way, if the presented analysis may have been influenced by a hypothetical high variability between splits.

Functions that select which subject to put in a specific set, operate under a random seed that can be set a priori.
Therefore, results are easily reproducible and can be used to verify the correct implementation of the N-LNSO strategy and check the variability between splits.
\autoref{tab: splitsummary} summarizes such information.
Ratios and percentile ranges refer to pipelines with the first and last segment removal, i.e., ICA and ICA+ASR. (Pipelines without segment removal, Raw and Filt, gave almost identical results, but their summary is not reported here for simplicity.)
As can be seen, variability is limited among all the 50 partitions per task.
The highest difference between the 5th and 95th percentile (still acceptable) is achieved by the test set class ratios of the Parkinson's task due to difficulties in aggregating two datasets with a very different size.

\renewcommand{\arraystretch}{2.5} % Default value: 1
\begin{table*}[!t]
%    \fontsize{8}{8}\selectfont
    \centering
    \caption{Leave N-Subject-Out splits summary. Median values are reported with the 5$^{th}$ and 95$^{th}$ percentile}
    \begin{tabular}{ccccccccc}
        \Xhline{2\arrayrulewidth} \\[-1,3em]
        \makecell{Task} & \makecell{Class \\ Label} & \makecell{Total \\ Class Ratio} & \makecell{Train \\ Ratio} & \makecell{Validation \\ Ratio} & \makecell{Test \\ Ratio} & \makecell{Train \\ Class Ratio} & \makecell{Validation \\ Class Ratio} & \makecell{Test \\ Class Ratio} \\
        \\[-1,3em] \Xhline{2\arrayrulewidth}
        
        \multicolumn{1}{c|}{\multirow{2}{*}{\makecell{Eye}}} &
        \multicolumn{1}{c|}{Eyes Closed} &
        \multicolumn{1}{c|}{0.500} &
        \multicolumn{1}{c|}{\multirow{2}{*}{\makecell{0.717 \\ \text{[0.717 - 0.733]}}}} & 
        \multicolumn{1}{c|}{\multirow{2}{*}{\makecell{0.183 \\ \text{[0.167 - 0.183]}}}} & 
        \multicolumn{1}{c|}{\multirow{2}{*}{\makecell{0.100 \\ \text{[0.100 - 0.100]}}}} & 
        \multicolumn{1}{c|}{\makecell{0.500 \\ \text{[0.500 - 0.500]}}} &
        \multicolumn{1}{c|}{\makecell{0.500 \\ \text{[0.500 - 0.500]}}} &
        \multicolumn{1}{c}{\makecell{0.500 \\ \text{[0.500 - 0.500]}}} \\
        \cline{2-3} \cline{7-9}
        \multicolumn{1}{c|}{} &
        \multicolumn{1}{c|}{Eyes Open} &
        \multicolumn{1}{c|}{\makecell{0.500}} &
        \multicolumn{1}{c|}{} &
        \multicolumn{1}{c|}{} &
        \multicolumn{1}{c|}{} &
        \multicolumn{1}{c|}{\makecell{0.500 \\ \text{[0.500 - 0.500]}}} &
        \multicolumn{1}{c|}{\makecell{0.500 \\ \text{[0.500 - 0.500]}}} &
        \multicolumn{1}{c}{\makecell{0.500 \\ \text{[0.500 - 0.500]}}} \\
        \hline

        \multicolumn{1}{c|}{\multirow{2}{*}{\makecell{Motor \\ Imagery}}} &
        \multicolumn{1}{c|}{Left Hand} &
        \multicolumn{1}{c|}{0.505} &
        \multicolumn{1}{c|}{\multirow{2}{*}{\makecell{0.718 \\ \text{[0.716 - 0.727]}}}} & 
        \multicolumn{1}{c|}{\multirow{2}{*}{\makecell{0.180 \\ \text{[0.178 - 0.188]}}}} & 
        \multicolumn{1}{c|}{\multirow{2}{*}{\makecell{0.103 \\ \text{[0.094 - 0.104]}}}} & 
        \multicolumn{1}{c|}{\makecell{0.504 \\ \text{[0.503 - 0.507]}}} &
        \multicolumn{1}{c|}{\makecell{0.505 \\ \text{[0.498 - 0.510]}}} &
        \multicolumn{1}{c}{\makecell{0.505 \\ \text{[0.497 - 0.511]}}} \\
        \cline{2-3} \cline{7-9}
        \multicolumn{1}{c|}{} &
        \multicolumn{1}{c|}{Right Hand} &
        \multicolumn{1}{c|}{\makecell{0.495}} &
        \multicolumn{1}{c|}{} &
        \multicolumn{1}{c|}{} &
        \multicolumn{1}{c|}{} &
        \multicolumn{1}{c|}{\makecell{0.496 \\ \text{[0.493 - 0.497]}}} &
        \multicolumn{1}{c|}{\makecell{0.495 \\ \text{[0.490 - 0.502]}}} &
        \multicolumn{1}{c}{\makecell{0.495 \\ \text{[0.489 - 0.503]}}} \\
        \hline

        \multicolumn{1}{c|}{\multirow{2}{*}{\makecell{Parkinson}}} &
        \multicolumn{1}{c|}{Control} &
        \multicolumn{1}{c|}{0.505} &
        \multicolumn{1}{c|}{\multirow{2}{*}{\makecell{0.720 \\ \text{[0.700 - 0.744]}}}} & 
        \multicolumn{1}{c|}{\multirow{2}{*}{\makecell{0.177 \\ \text{[0.157 - 0.204]}}}} & 
        \multicolumn{1}{c|}{\multirow{2}{*}{\makecell{0.099 \\ \text{[0.092 - 0.110]}}}} & 
        \multicolumn{1}{c|}{\makecell{0.507 \\ \text{[0.473 - 0.528]}}} &
        \multicolumn{1}{c|}{\makecell{0.492 \\ \text{[0.433 - 0.568]}}} &
        \multicolumn{1}{c}{\makecell{0.488 \\ \text{[0.359 - 0.666]}}} \\
        \cline{2-3} \cline{7-9}
        \multicolumn{1}{c|}{} &
        \multicolumn{1}{c|}{Parkinson} &
        \multicolumn{1}{c|}{\makecell{0.495}} &
        \multicolumn{1}{c|}{} &
        \multicolumn{1}{c|}{} &
        \multicolumn{1}{c|}{} &
        \multicolumn{1}{c|}{\makecell{0.493 \\ \text{[0.472 - 0.527]}}} &
        \multicolumn{1}{c|}{\makecell{0.508 \\ \text{[0.432 - 0.567]}}} &
        \multicolumn{1}{c}{\makecell{0.512 \\ \text{[0.334 - 0.641]}}} \\
        \hline

        \multicolumn{1}{c|}{\multirow{3}{*}{Alzheimer}} &
        \multicolumn{1}{c|}{Control} &
        \multicolumn{1}{c|}{0.347} &
        \multicolumn{1}{c|}{\multirow{3}{*}{\makecell{0.717 \\ \text{[0.699 - 0.745]}}}} & 
        \multicolumn{1}{c|}{\multirow{3}{*}{\makecell{0.183 \\ \text{[0.158 - 0.196]}}}} & 
        \multicolumn{1}{c|}{\multirow{3}{*}{\makecell{0.102 \\ \text{[0.078 - 0.117]}}}} & 
        \multicolumn{1}{c|}{\makecell{0.347 \\ \text{[0.331 - 0.359]}}} &
        \multicolumn{1}{c|}{\makecell{0.342 \\ \text{[0.307 - 0.43]}}} &
        \multicolumn{1}{c}{\makecell{0.347 \\ \text{[0.304 - 0.384]}}} \\
        \cline{2-3} \cline{7-9}
        \multicolumn{1}{c|}{} &
        \multicolumn{1}{c|}{\makecell{Frontotemporal \\ Dementia}} &
        \multicolumn{1}{c|}{0.237} &
        \multicolumn{1}{c|}{} &
        \multicolumn{1}{c|}{} &
        \multicolumn{1}{c|}{} &
        \multicolumn{1}{c|}{\makecell{0.237 \\ \text{[0.224 - 0.246]}}} &
        \multicolumn{1}{c|}{\makecell{0.239 \\ \text{[0.195 - 0.266]}}} &
        \multicolumn{1}{c}{\makecell{0.239 \\ \text{[0.162 - 0.324]}}} \\
        \cline{2-3} \cline{7-9}
        \multicolumn{1}{c|}{} &
        \multicolumn{1}{c|}{Alzheimer} &
        \multicolumn{1}{c|}{0.417} &
        \multicolumn{1}{c|}{} &
        \multicolumn{1}{c|}{} &
        \multicolumn{1}{c|}{} &
        \multicolumn{1}{c|}{\makecell{0.416 \\ \text{[0.399 - 0.433]}}} &
        \multicolumn{1}{c|}{\makecell{0.427 \\ \text{[0.358 - 0.462]}}} &
        \multicolumn{1}{c}{\makecell{0.428 \\ \text{[0.315 - 0.485]}}} \\
        \hline

        \multicolumn{1}{c|}{\multirow{2}{*}{\makecell{Sleep \\ Deprivation}}} &
        \multicolumn{1}{c|}{Control} &
        \multicolumn{1}{c|}{0.500} &
        \multicolumn{1}{c|}{\multirow{2}{*}{\makecell{0.717 \\ \text{[0.707 - 0.731]}}}} & 
        \multicolumn{1}{c|}{\multirow{2}{*}{\makecell{0.185 \\ \text{[0.170 - 0.185]}}}} & 
        \multicolumn{1}{c|}{\multirow{2}{*}{\makecell{0.099 \\ \text{[0.090 - 0.114]}}}} & 
        \multicolumn{1}{c|}{\makecell{0.500 \\ \text{[0.498 - 0.503]}}} &
        \multicolumn{1}{c|}{\makecell{0.503 \\ \text{[0.488 - 0.505]}}} &
        \multicolumn{1}{c}{\makecell{0.500 \\ \text{[0.480 - 0.508]}}} \\
        \cline{2-3} \cline{7-9}
        \multicolumn{1}{c|}{} &
        \multicolumn{1}{c|}{\makecell{Sleep \\ Deprived}} &
        \multicolumn{1}{c|}{\makecell{0.500}} &
        \multicolumn{1}{c|}{} &
        \multicolumn{1}{c|}{} &
        \multicolumn{1}{c|}{} &
        \multicolumn{1}{c|}{\makecell{0.500 \\ \text{[0.497 - 0.502]}}} &
        \multicolumn{1}{c|}{\makecell{0.497 \\ \text{[0.495 - 0.512]}}} &
        \multicolumn{1}{c}{\makecell{0.500 \\ \text{[0.492 - 0.520]}}} \\
        \hline
        
        \multicolumn{1}{c|}{\multirow{2}{*}{\makecell{First \\ Episode \\ Psychosis}}} &
        \multicolumn{1}{c|}{Control} &
        \multicolumn{1}{c|}{0.486} &
        \multicolumn{1}{c|}{\multirow{2}{*}{\makecell{0.721 \\ \text{[0.710 - 0.726]}}}} & 
        \multicolumn{1}{c|}{\multirow{2}{*}{\makecell{0.179 \\ \text{[0.177 - 0.186]}}}} & 
        \multicolumn{1}{c|}{\multirow{2}{*}{\makecell{0.098 \\ \text{[0.096 - 0.112]}}}} & 
        \multicolumn{1}{c|}{\makecell{0.492 \\ \text{[0.469 - 0.499]}}} &
        \multicolumn{1}{c|}{\makecell{0.456 \\ \text{[0.442 - 0.551]}}} &
        \multicolumn{1}{c}{\makecell{0.493 \\ \text{[0.425 - 0.508]}}} \\
        \cline{2-3} \cline{7-9}
        \multicolumn{1}{c|}{} &
        \multicolumn{1}{c|}{FEP} &
        \multicolumn{1}{c|}{\makecell{0.514}} &
        \multicolumn{1}{c|}{} &
        \multicolumn{1}{c|}{} &
        \multicolumn{1}{c|}{} &
        \multicolumn{1}{c|}{\makecell{0.508 \\ \text{[0.501 - 0.531]}}} &
        \multicolumn{1}{c|}{\makecell{0.554 \\ \text{[0.449 - 0.558]}}} &
        \multicolumn{1}{c}{\makecell{0.507 \\ \text{[0.492 - 0.575]}}} \\
        \hline
    \end{tabular}
    \label{tab: splitsummary}
\end{table*}
\renewcommand{\arraystretch}{1} % Default value: 1

\subsubsection{Why splitting at the subject level is important?}

\begin{figure}[!t]
    \centering
    \includegraphics[width=0.85\linewidth]{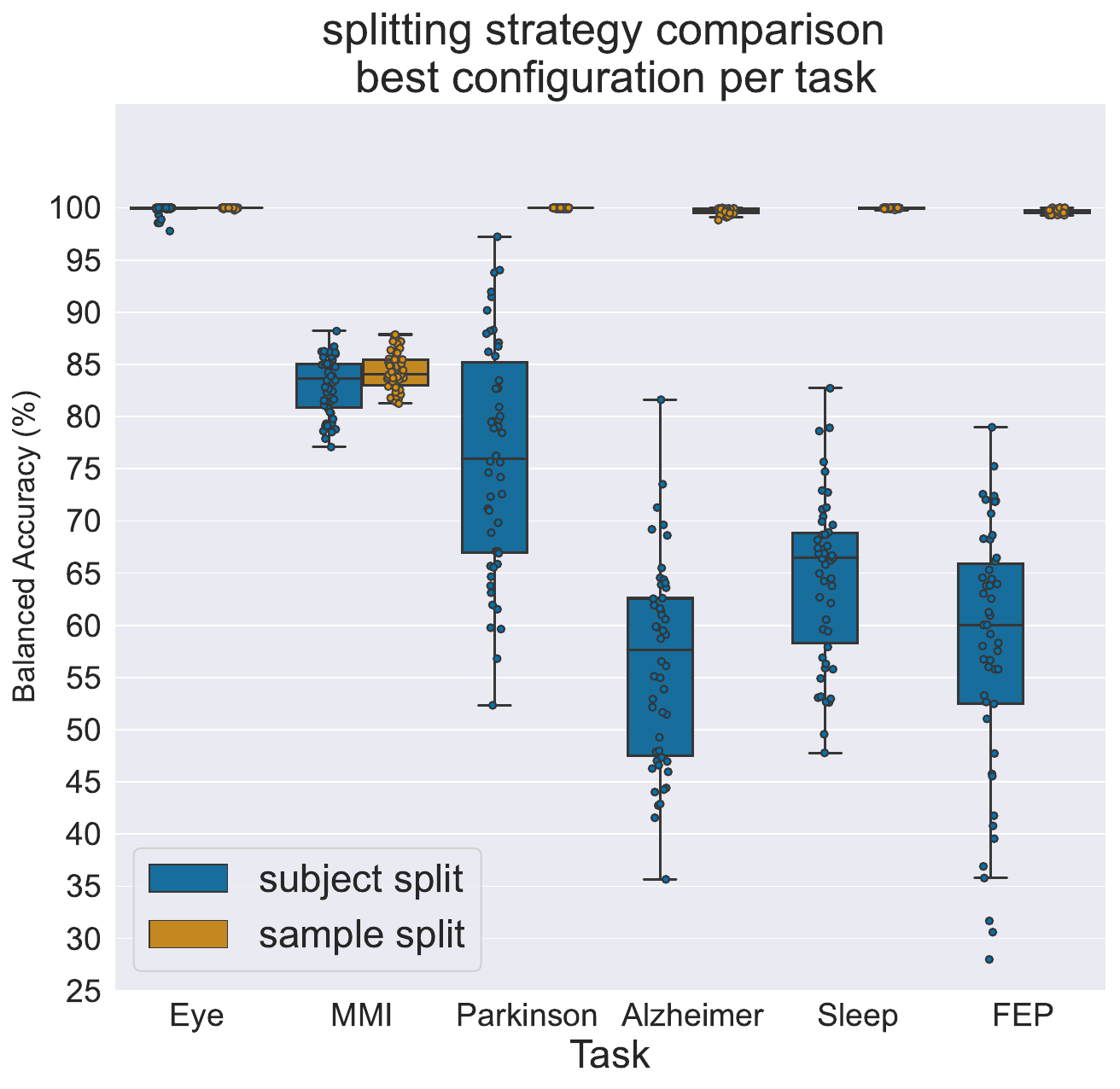}
    \caption{
    Comparison of results between a subject level splitting strategy and a sample level one. 
    Subject split results refer to those achieved with the best configuration (model + pipeline) when using the Nested Leave-N-Subject-Out method.
    Sample split results refer to those achieved with the same best configuration but using a splitting procedure at the sample level, i.e., by assigning samples from the same subject to both training, validation, and test sets.
    }
    \label{fig: sample_split}
\end{figure}

As presented in the ``Data Partition" section of the paper, several practical reasons bring to the choice of an inter-subject evaluation, proposed in the form of a variant of the Leave-N-Subject-Out (LNSO).
First of all, most of the real-world applications expect models to be used on new subjects rather than on new samples of the same ones.
Secondly, it is known that intra-subject evaluations could end up producing models whose embeddings are dominated by subject-specific characteristics that could lead to misleading results: extremely good on subjects already seen but rather poor on unseen ones.
However, one might ask how results can differ between these two strategies.

To answer this question, an additional set of training with a hold out partition strategy at the sample level was performed.
In particular, each split was created by firstly selecting a random portion of consecutive windows (samples) from each dataset's EEG record, which were included in the test set; then, by allocating the remaining samples in the training and validation sets in such a way that no adjacent windows can be found between the training and test set.
This split was repeated 50 times with different seed $$\text{seed}_i = \lfloor 83136297/i\rfloor$$ with $i\in[1, 50]$ in order to replicate the same number of partitions of the N-LNSO procedure proposed in the paper.

\autoref{fig: sample_split} shows how different results are when changing the splitting strategy from a subject to a sample level. 
Subject split results refer to those achieved with the best configuration (model + pipeline) when using the Nested Leave-N-Subject-Out method, while sample split results refer to those achieved with the same best configuration but using the sample-based splitting procedure described above.

As can be seen, if the model is allowed to see a portion of EEG from each subject, its prediction are most of the times almost perfect when seeing another portion of the same record.
This would have created unrealistic results both in terms of model capabilities and rank between different preprocessing pipelines. 

%\clearpage
\subsection{Do the results change when using another metric?}
\label{SM-metrics}

This section shows how different evaluation metrics affect the significance of the Friedman's test $\chi^2_F$; results are just reported as significant $(*)$ if the statistic $\chi^2_F\geq7.6$, which is the tabulated Friedman' statistic for $k=4,N=6$.
As reported in \autoref{tab: accuracies}, except for precision weighted, the significance is found only in DeepConvNet for all the other metrics considered.

\renewcommand{\arraystretch}{1.6} % Default value: 1
\begin{table}[h]
    \centering
    \caption{Friedman's test $\chi^2_F$ values for different metrics.}
    \begin{tabular}{ccccc}
        \Xhline{2\arrayrulewidth} \\[-1,3em]
        Metric & EEGNet & ShallowNet & DeepConvNet & FBCNet \\
        \\[-1,3em] \Xhline{2\arrayrulewidth}
        \multicolumn{1}{c|}{\makecell{Unbalanced \\ accuracy}} & \multicolumn{1}{c|}{3.0} & \multicolumn{1}{c|}{3.8} & \multicolumn{1}{c|}{11.4 $(*)$} & 4.2 \\
        \hline
        \multicolumn{1}{c|}{\makecell{Balanced \\ accuracy}} & \multicolumn{1}{c|}{1.8} & \multicolumn{1}{c|}{4.2} & \multicolumn{1}{c|}{8.8 $(*)$} & 3.8 \\
        \hline
        \multicolumn{1}{c|}{\makecell{F1-score \\ weighted}} & \multicolumn{1}{c|}{3.8} & \multicolumn{1}{c|}{4.2} & \multicolumn{1}{c|}{12.2 $(*)$} & 4.2 \\
        \hline
        \multicolumn{1}{c|}{\makecell{Precision \\ weighted}} & \multicolumn{1}{c|}{4.6} & \multicolumn{1}{c|}{4.2} & \multicolumn{1}{c|}{7.4} & 4.2 \\
        \hline
        \multicolumn{1}{c|}{\makecell{Recall \\ weighted}} & \multicolumn{1}{c|}{3.0} & \multicolumn{1}{c|}{3.8} & \multicolumn{1}{c|}{11.4 $(*)$} & 4.2 \\
        \hline
        \multicolumn{1}{c|}{\makecell{Cohen's \\Kappa}} & \multicolumn{1}{c|}{3.0} & \multicolumn{1}{c|}{4.2} & \multicolumn{1}{c|}{11.0 $(*)$} & 4.2 \\        
        \hline
    \end{tabular}
    \label{tab: accuracies}
\end{table}
\renewcommand{\arraystretch}{1} % Default value: 1

\subsection{Task marginalization}
\label{SM-taskmargin}

\begin{figure*}[!t]
    \centering
    \includegraphics[width=0.47\textwidth]{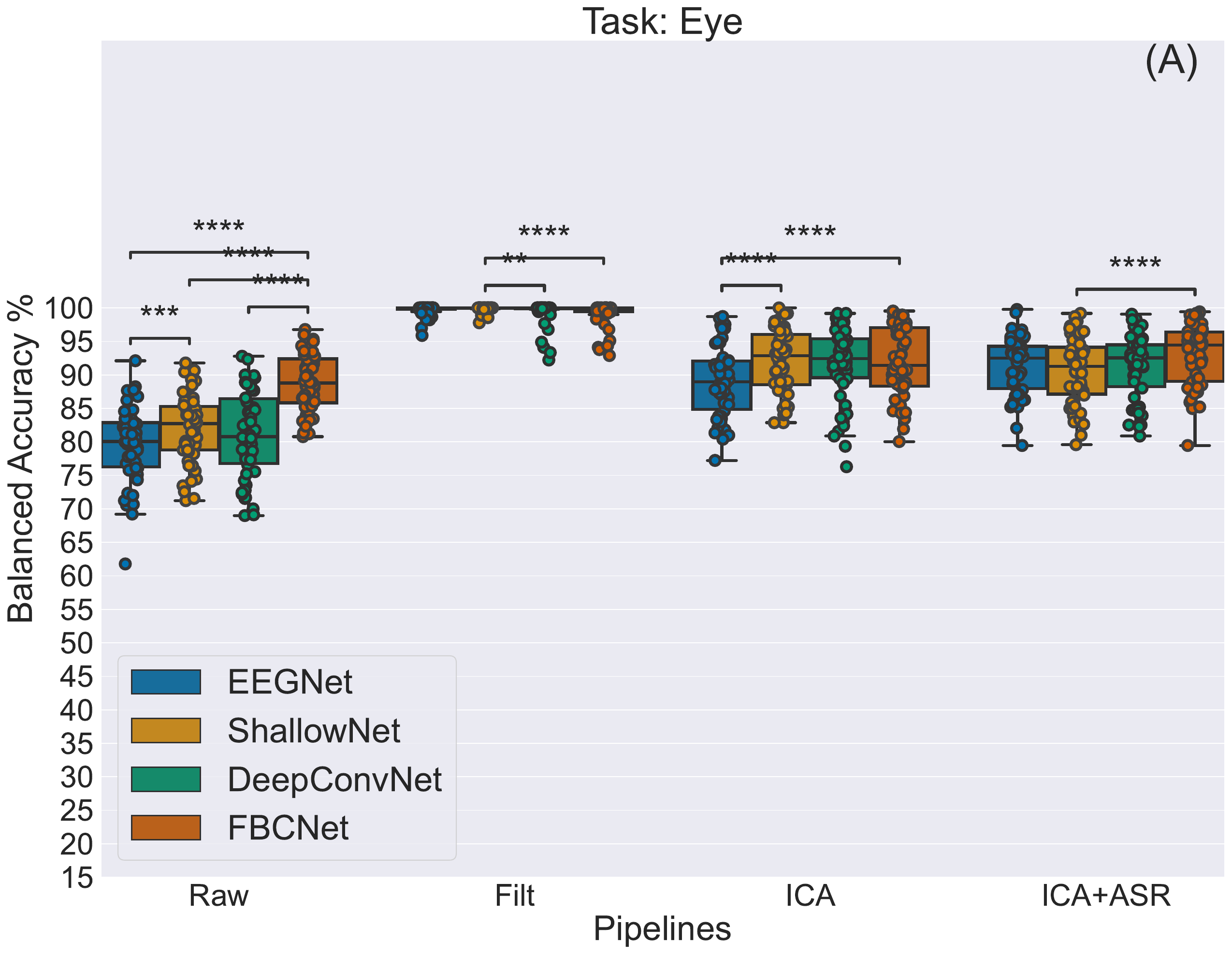}
    \includegraphics[width=0.47\textwidth]{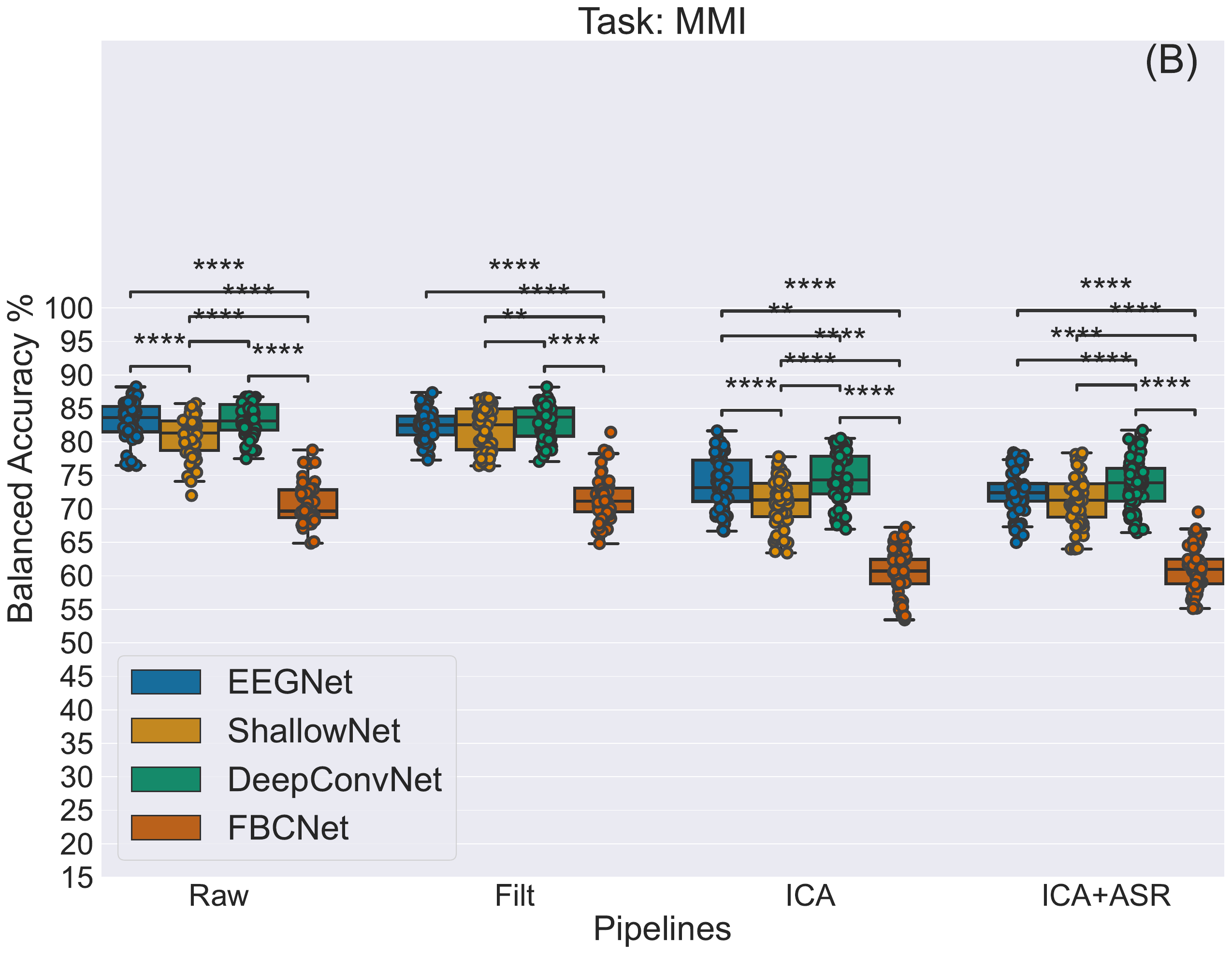}
    
    \includegraphics[width=0.47\textwidth]{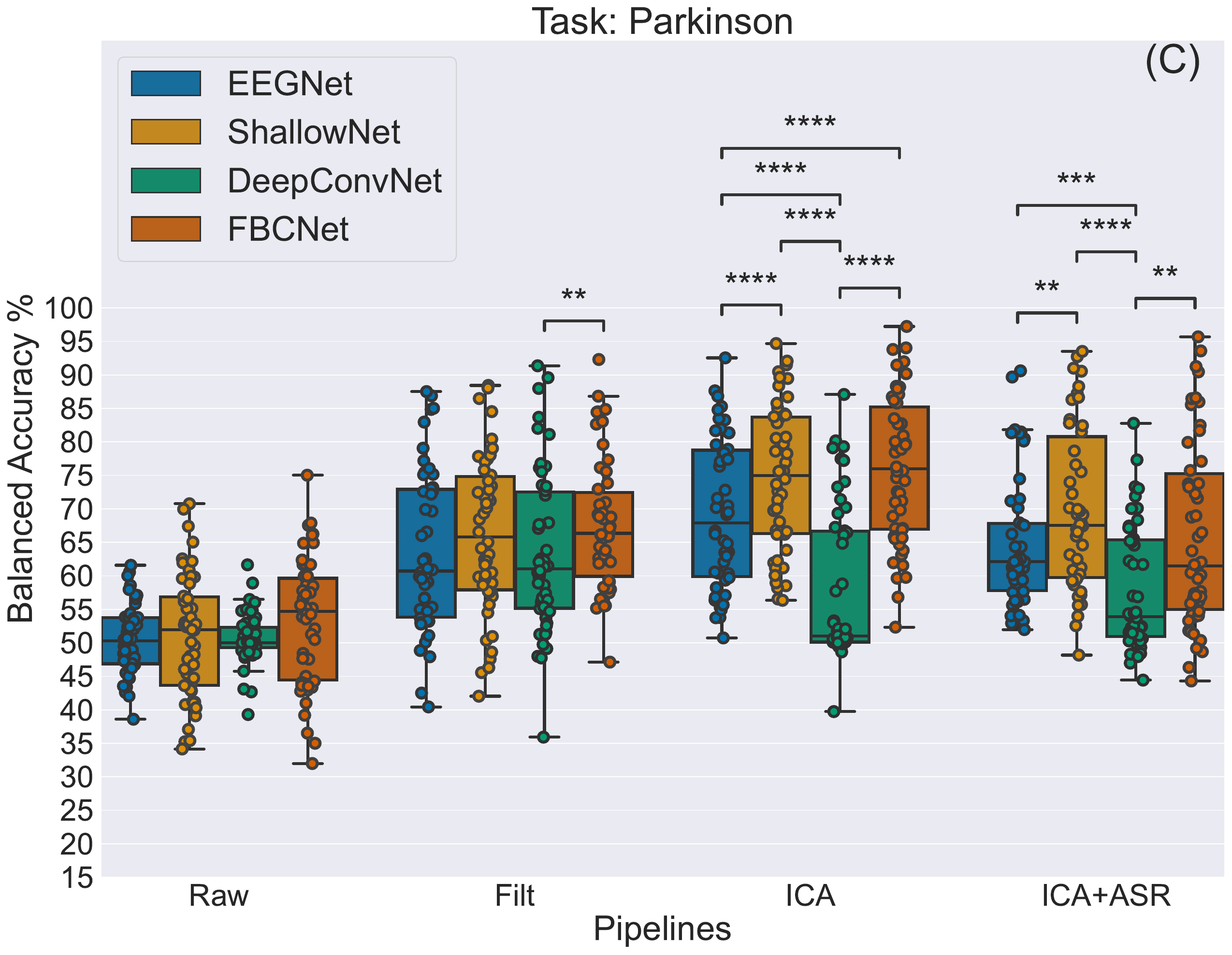}
    \includegraphics[width=0.47\textwidth]{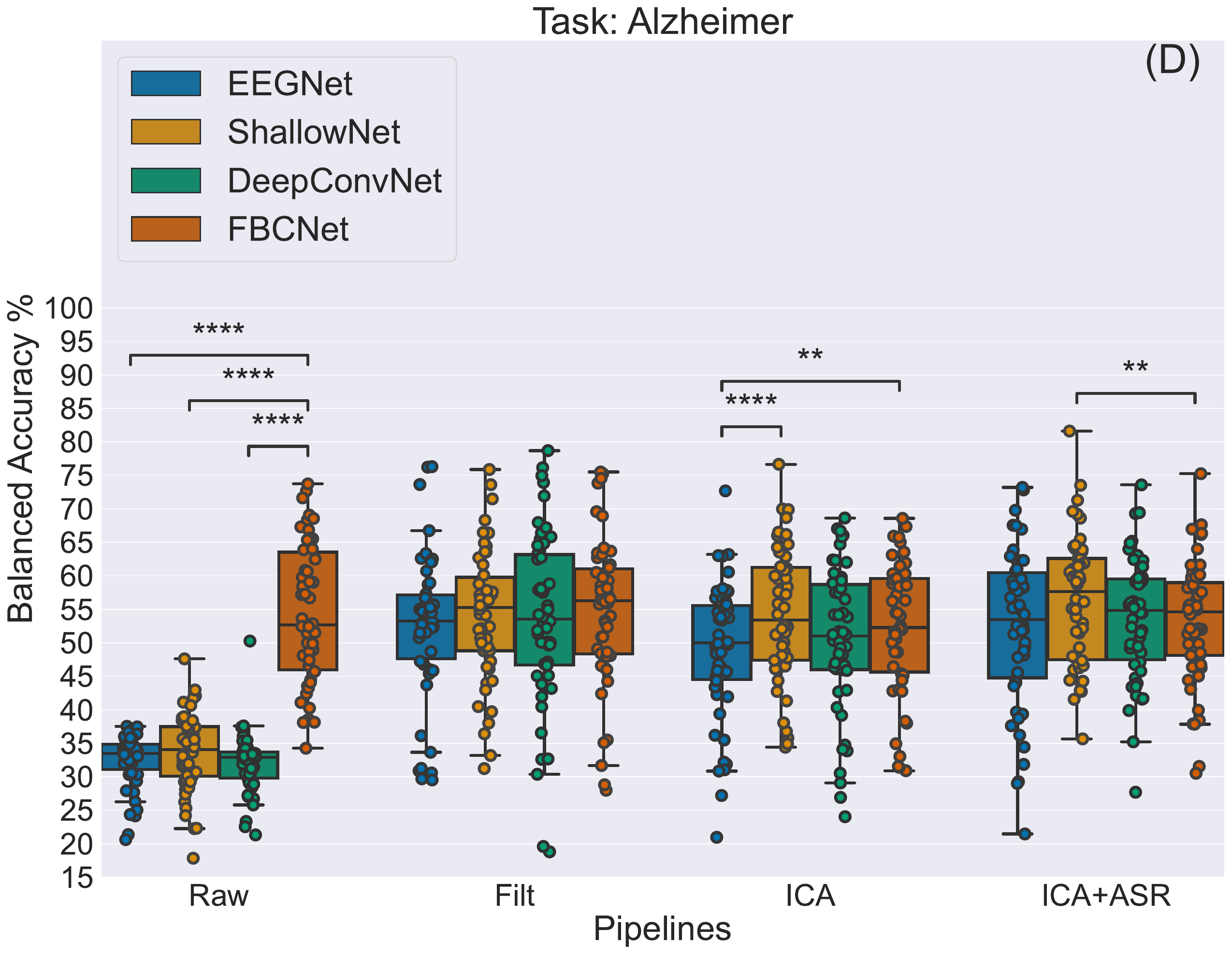}
    
    \includegraphics[width=0.47\textwidth]{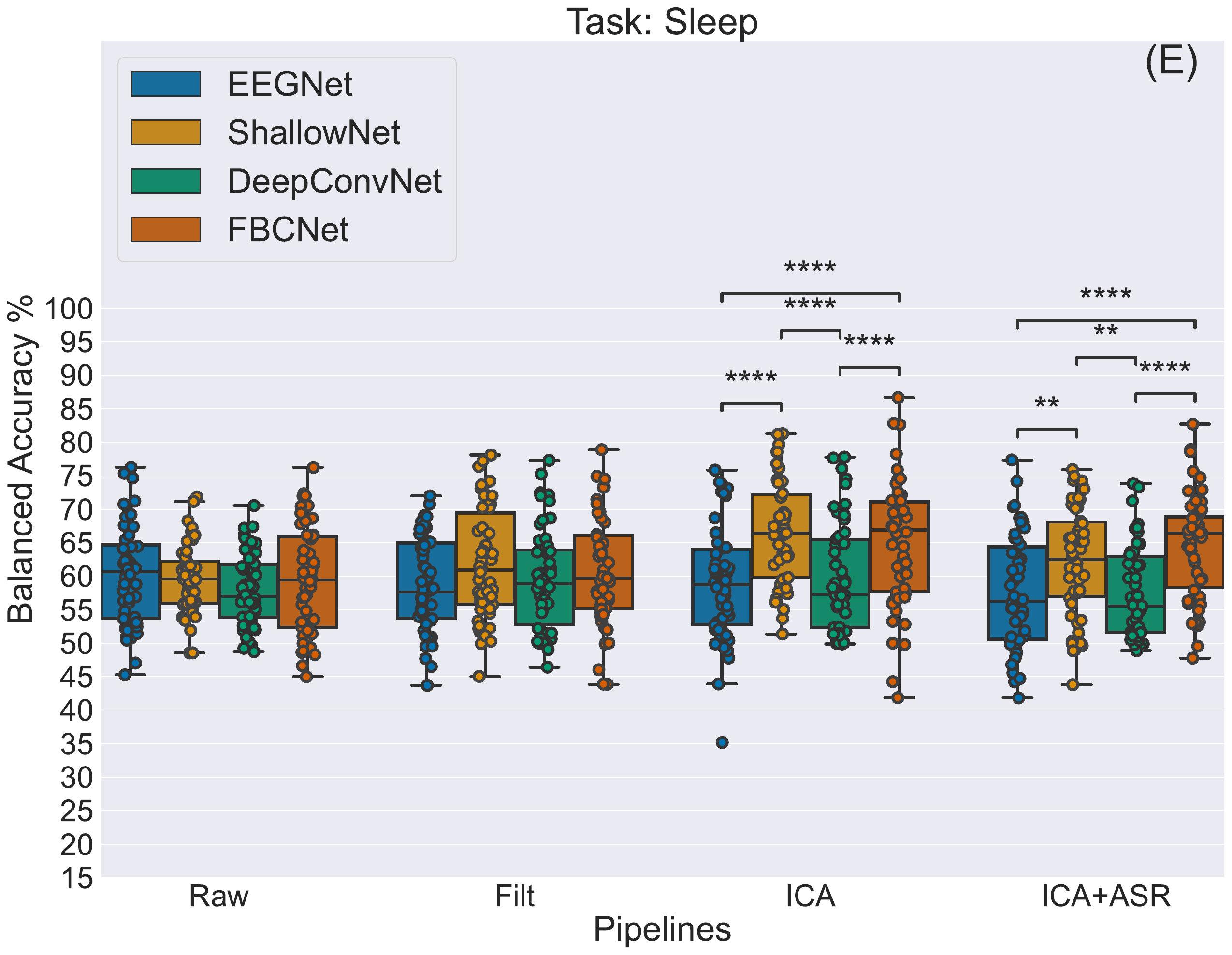}
    \includegraphics[width=0.47\textwidth]{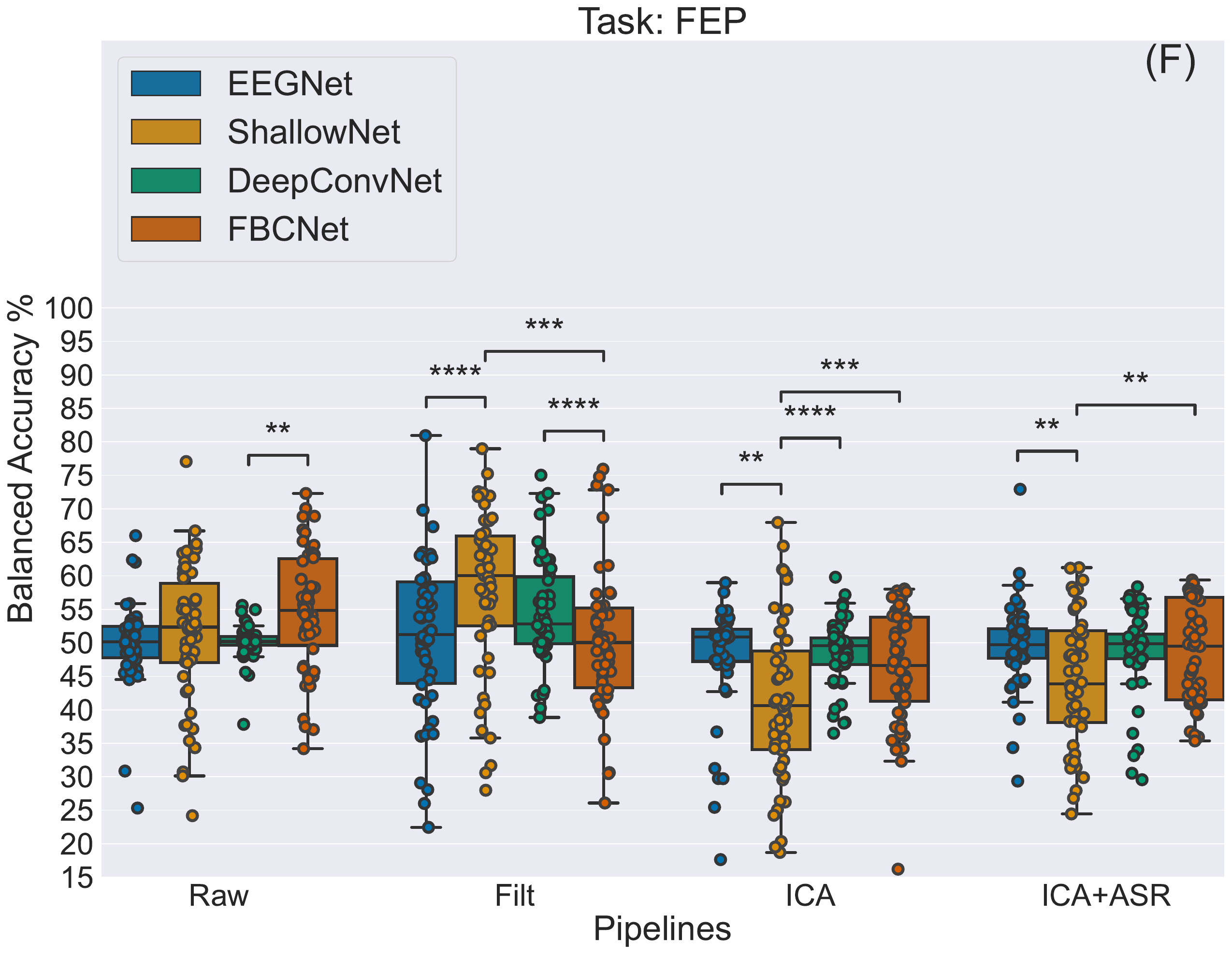}
    
    \caption{Panels A to F shows the balanced accuracies for the 4 models and for the 4 pipelines investigated.
    Results are marginalized by tasks, which are: in Panel A, Eye, in Panel B MMI, in Panel C Parkinson, in Panel D Alzheimer, in Panel E Sleep and in Panel F FEP.
    Asterisks indicates significant results from Wilcoxon's signed rank tests, performed within the same pipeline: ($**$$**$) $p<0.0001$; ($**$$*$) $p<0.001$; ($**$) $p<0.01$ and ($*$) $p<0.05$.
    Results are Holm corrected, for multiple comparison within the same pipeline.}
    \label{fig: task_results}
\end{figure*}

This section provides an alternative visualization of the results presented in Section III, Figure 2, of the main body.
While the main objective of the article is evaluating the effect of preprocessing using different pipelines, using the six tasks as evidence, in \autoref{fig: task_results} results are instead marginalized by task.
This visualization helps those who want to perform one of the presented tasks with deep learning strategies.
If the preprocessing pipeline is decided a priori, then the reader can see which architecture gives the best performance.
Instead, if the architecture is decided a priori, then the reader can see which preprocessing pipeline gives the best performance.

%\begin{IEEEbiography}[{\includegraphics[width=1in,height=1.25in,clip,keepaspectratio]{a1.png}}]{First A. Author} photograph and biography available at the time of publication
%\end{IEEEbiography}

%\begin{IEEEbiographynophoto}{First author} photograph and biography not available at the
%time of publication.

\end{document}